\documentclass{iopart}
\usepackage{graphicx}
\usepackage{latexsym}
\usepackage{amssymb}

\jl{6}        
\eqnobysec    

\font\fiverm=cmr5

\input prepictex
\input pictex
\input postpictex

\def\mathput#1{\relax \ifmmode \displaystyle #1\else $\displaystyle #1$\fi}

\def\beq{\begin{equation}}
\def\eeq{\end{equation}}
\def\rmd{{\rm d}}

\begin{document}

\title[The Erez-Rosen metric and the role of the quadrupole on light propagation]
{The Erez-Rosen metric and the role of the quadrupole on light propagation}

\author{
Donato Bini$^{a,b,c,d}$,
Mariateresa Crosta${}^{d}$,
Fernando de Felice${}^{e}$,
Andrea Geralico${}^{b,f}$ and
Alberto Vecchiato${}^{d}$
}

\address{${}^a$\
Istituto per le Applicazioni del Calcolo ``M. Picone,'' CNR,
I--00185 Rome, Italy
}

\address{${}^b$\
ICRA, International Center for Relativistic Astrophysics, University of Rome "La Sapienza,"
I--00185 Rome, Italy
}

\address{${}^c$\
INFN, Sezione di Firenze, Polo Scientifico, Via Sansone 1, 
I--50019 Sesto Fiorentino, Florence, Italy 
}

\address{${}^d$\
INAF, Astronomical Observatory of Torino
via Osservatorio 20, I--10025 Pino Torinese (TO), Italy
}

\address{${}^e$\
Department of Physics and Astronomy, University of Padova,
via Marzolo 8, I--35131 Padova, Italy
}

\address{
  $^{f}$
  Physics Department,
  University of Rome ``La Sapienza,'' I--00185 Rome, Italy
}

\begin{abstract}
The gravitational field of a static body with quadrupole moment is described by an exact solution found by Erez and Rosen. Here we investigate the role of the quadrupole in the motion, deflection and lensing of a light ray in the above metric. The standard lensing observables like image positions and magnification have been explicitly obtained in the weak field and small quadrupole limit.
In this limit the spacetime metric appears as the natural generalization to
quadrupole corrections of the metric form adopted also in current astrometric models.
Hence, the corresponding analytical solution of the inverse ray tracing problem as well
as the consistency with other approaches are also discussed.
\end{abstract}

\pacno{04.20.Cv}

\section{Introduction}

Nowadays, modern space technology allows for high accuracies in astronomical  observations.
Attaining high level of precision implies that the effects on the propagation of surface skimming photons
due to the quadrupole moment of  sources in the Solar System (like Jupiter, see, e.g., Refs. \cite{cromi,kop3} and references therein)  
are no longer negligible, since they contribute to the above mentioned order.
When approached in its full generality, however, the problem of accounting for quadrupole corrections (and higher order polarities) cannot be solved exactly, since it turns out to be very difficult from both a physical and a mathematical point of view. Even the simplest case of a single gravitating body endowed  with a quadrupolar structure is
barely tractable.
As a consequence, several approximations have been proposed as the only way to get an explicit solution and different approaches are now available in the literature (see, e.g., Ref. \cite{willbook}).

However, an exact solution of the vacuum Einstein equations describing the spacetime of a quadrupolar body exists and is given by the Erez-Rosen solution \cite{erez}.
This has been extensively studied mainly in the strong gravitational regime, namely in the vicinity of the metric source.
In the present paper we extend the analysis to the weak field limit having in mind  that can be applied to general relativistic astrometry, 
in particular to the next generation of high accurate astrometric missions like Gaia (ESA) \cite{turon} that will almost ready to be launched in October 2013.

We first concentrate on certain optical properties of the Erez-Rosen
solution extending previous investigations \cite{quev90,masquev95}, then we compare and
contrast among approximate and exact one-body solutions, using ray tracing
as a scouting device and analyzing in
a fully analytical scheme how the quadrupole moment modifies all standard
formulas
involving light propagation, deflection and lensing.

We find that ray tracing in the weak field and small quadrupole limit of the Erez-Rosen metric is fully consistent with the analysis made {\it ab initio} in a post-Newtonian approximation with quadrupole corrections added, highlighting the complexity of the quadrupole perturbations.

Hereafter, latin indices run from 1 to 3, greek indices from 0 to 3.

\section{An exact vacuum solution for a quasi-spherical source: the Erez-Rosen metric}

The metric of a nonrotating mass with a quadrupole moment has been obtained  by Erez and Rosen \cite{erez}, later corrected for several numerical coefficients  by Doroshkevich et al. \cite{novikov} and Young and Coulter \cite{young}.
It belongs to the static Weyl class of solutions with the line element written in prolate spheroidal coordinates ($t,x,y,\phi$), with $x \geq 1$ and $-1 \leq y \leq 1$, as follows \cite{ES}
\beq\fl\quad
\label{metric_Weyl}
\rmd s^2=-f\rmd t^2+\frac{\sigma^2}{f}\left\{e^{2\gamma}\left(x^2-y^2\right)\left(\frac{\rmd x^2}{x^2-1}+\frac{\rmd y^2}{1-y^2}\right)+(x^2-1)(1-y^2)\rmd\phi^2\right\}\,,
\eeq
where $\sigma$ is a constant and the quantities $f$ and $\gamma$ are functions of $x$ and $y$ only.
The metric functions are given by
\begin{eqnarray}\fl\quad
\label{metdef}
f&=&\frac{x-1}{x+1}e^{-2qP_2Q_2}\,, \nonumber\\
\fl\quad
\gamma&=&\frac12(1+q)^2 \ln\frac{x^2-1}{x^2-y^2} + 2q(1-P_2)Q_1 + q^2(1-P_2) \bigg[ (1+P_2)(Q_1^2-Q_2^2)\nonumber \\
\fl\quad
&&+\frac12(x^2-1)(2Q_2^2 - 3xQ_1Q_2 + 3 Q_0Q_2 - Q_2')\bigg] \,.
\end{eqnarray}
Here $P_l(y)$ and $Q_l(x)$ are Legendre polynomials of the first and second kind, respectively, and $q$ is the dimensionless quadrupolar parameter. 
When $q=0$, the metric (\ref{metric_Weyl}) reduces to the Schwarzschild solution provided we identify $\sigma$ with the mass of the source, namely $\sigma=M$.
Transition of this  metric form to the more familiar one associated with standard Schwarzschild-like coordinates is accomplished by the following coordinate transformation
\beq
\label{trasftoBL}
x=\frac{r}{M}-1\,, \qquad
y=\cos\theta\,.
\eeq
According to the Geroch-Hansen \cite{ger,hans} definition of relativistic multipole moments ${\mathcal M}_n$, the mass monopole moment associated with this solution is ${\mathcal M}_0=M$ and the quadrupole moment is given by ${\mathcal M}_2=(2/15)qM^3$.
Negative values of the parameter $q$ correspond to oblate configurations, whereas positive values to prolate ones, relative to the axis $y=\pm1$.
Higher multipole moments of the order $n=2k$, $k=2,3,4,\ldots$, are determined by $q$ and $M$ in such a way that they all vanish when $q=0$.

The presence of the quadrupole parameter $q$ changes significantly the structure of the spacetime as compared with the Schwarzschild solution. In particular the hypersurface $x=1$, which is null in the Schwarzschild case, becomes directionally singular, its properties depending both on the value of $q$ and, as stated, the direction of approach.
For instance, in the equatorial plane this hypersurface is null for $(q-1)^2<5$ and timelike otherwise.
We refer to Ref. \cite{quev90} and to the Appendix A of Ref. \cite{masquev95} for a careful analysis of the causality properties associated with constant coordinate time slicings of the Erez-Rosen metric (\ref{metric_Weyl}).

Indeed, there are new geometrical properties which are worth to be explored. Let us consider for instance the spectral type of the solution which is algebraically general.
Defining the complex tensor
\beq
\tilde C_{\alpha\beta\gamma\delta}=C_{\alpha\beta\gamma\delta}-i{}^*C_{\alpha\beta\gamma\delta}\,,
\eeq
where $C_{\alpha\beta\gamma\delta}$ is the Weyl tensor and $*C_{\alpha\beta\gamma\delta}$ its left-dual,
one can introduce the two curvature invariants
\begin{equation}
I=\frac 1{32}\tilde{C}_{\alpha\beta\gamma\delta}\tilde{C}^{\alpha\beta\gamma\delta}
\,,\qquad
J=\frac 1{384}\tilde{C}_{\alpha\beta\gamma\delta}\tilde{C}^{\gamma\delta}{}_{\mu\nu}\tilde{C}^{\mu\nu\alpha\beta}\,.
\end{equation}
They can be used to define the speciality index~\cite{BC,beetle}
\begin{equation}
\label{SPECT}
\mathcal{S}=\frac{27J^2}{I^3}\,,
\end{equation}
whose value demarcates, in an invariant way, the transition from algebraically special solutions ($\mathcal{S}=1$) to solutions of  Petrov type-I ($\mathcal{S}\neq 1$)~\cite{ES}.
In the case of the Erez-Rosen metric, $I$,  $J$ and  $\mathcal{S}$ turn out to be  real.
It has been shown in Ref. \cite{quev90} that the Erez-Rosen solution is everywhere of Petrov type-I, except on the symmetry axis $y=\pm1$, where it is of type-D.

\begin{figure}
\typeout{*** EPS figure 2}
\begin{center}
\includegraphics[scale=0.35]{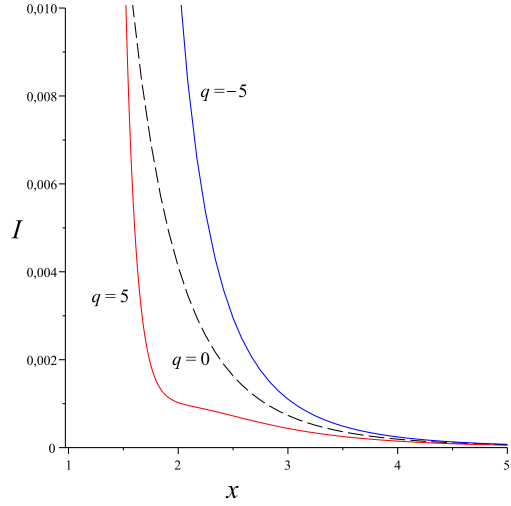}
\end{center}
\caption{The invariant $I$ as a function of $x$ for $y=0$ and $q=[-5,0,5]$.
}
\label{fig:Iqvari}
\end{figure}

\begin{figure}
\typeout{*** EPS figure 2}
\begin{center}
$\begin{array}{cc}
\includegraphics[scale=0.3]{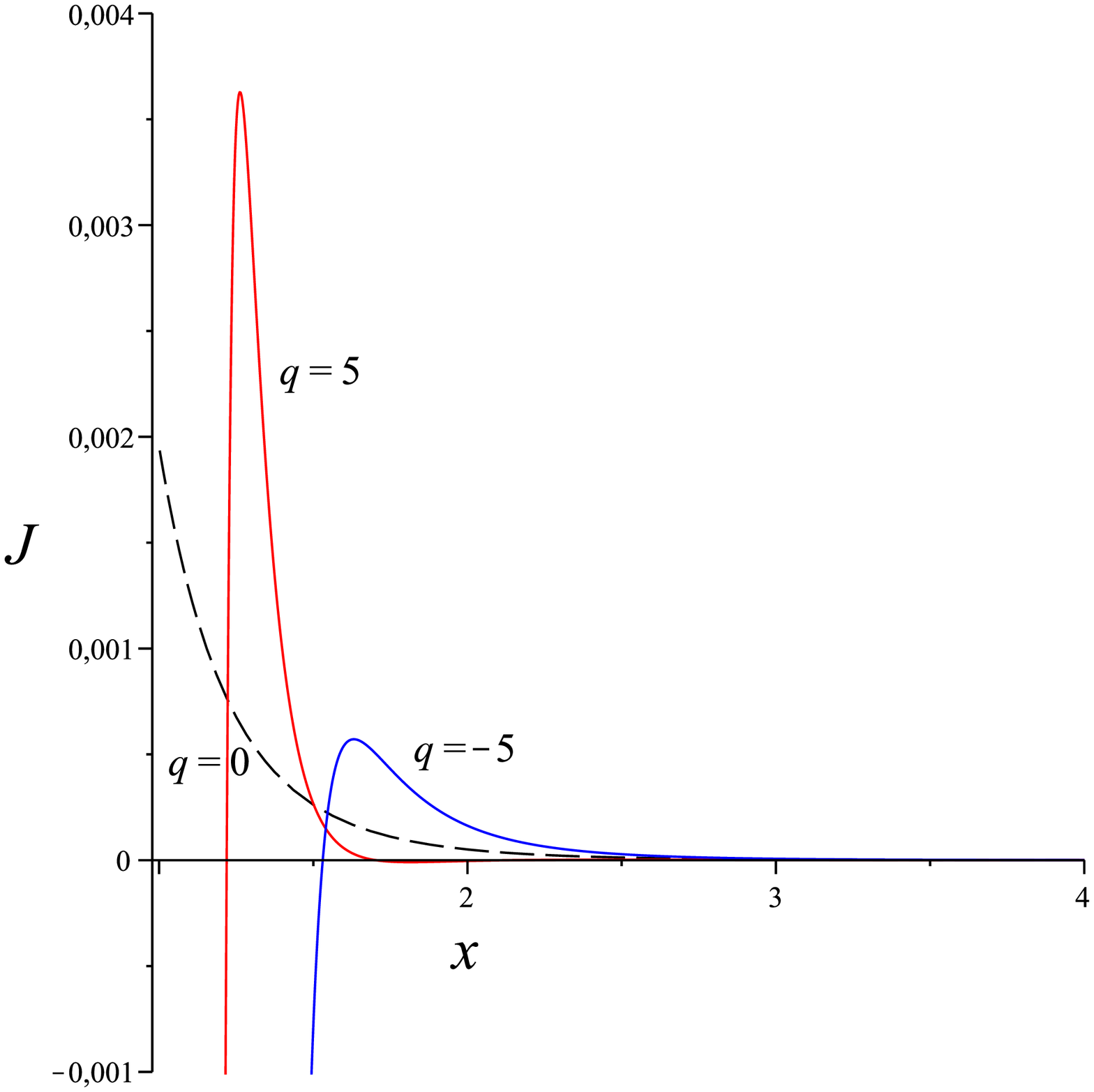}&\quad
\includegraphics[scale=0.3]{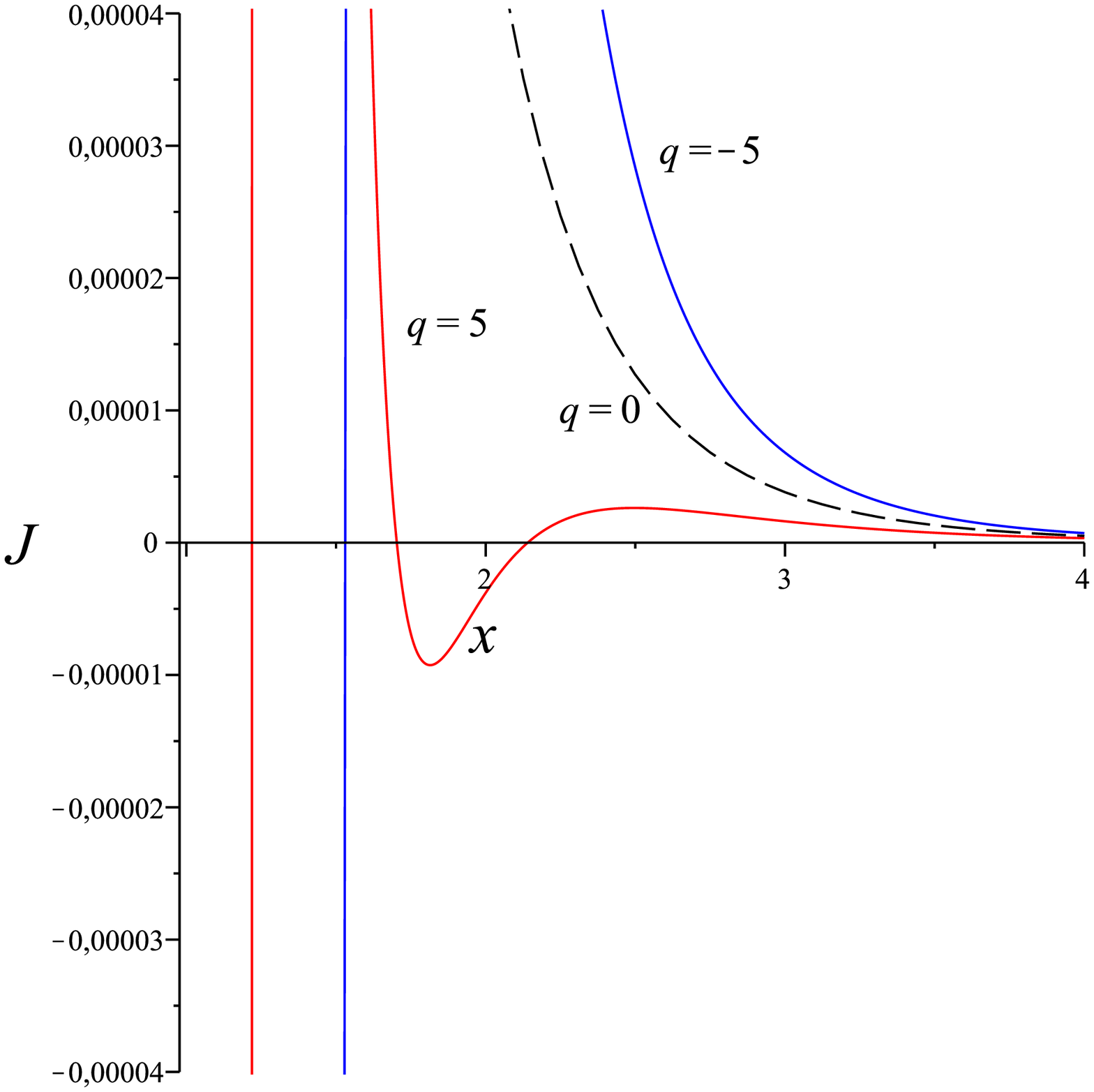}\\[.4cm]
\quad\mbox{(a)}\quad &\quad \mbox{(b)}
\end{array}$\\
\end{center}
\caption{The invariant $J$ as a function of $x$ for $y=0$ and $q=[-5,0,5]$.
In contrast to the Schwarzschild case, the presence of the quadrupole moment causes $J$ to change its sign close to the singularity (see also the closeup in panel (b)).
}
\label{fig:Jqvari}
\end{figure}

\begin{figure}
\typeout{*** EPS figure 1}
\begin{center}
$\begin{array}{cc}
\includegraphics[scale=0.3]{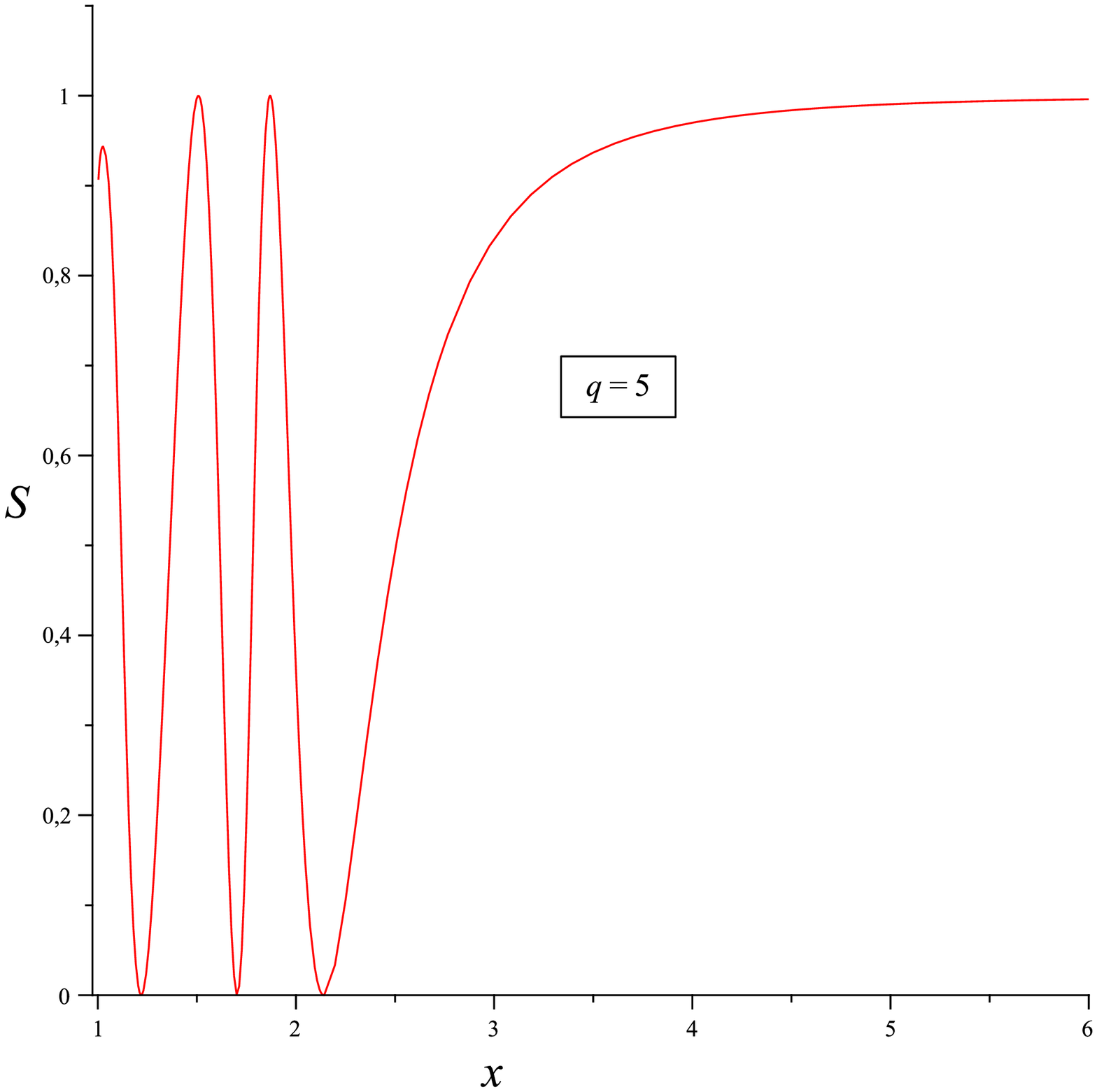}&\quad
\includegraphics[scale=0.3]{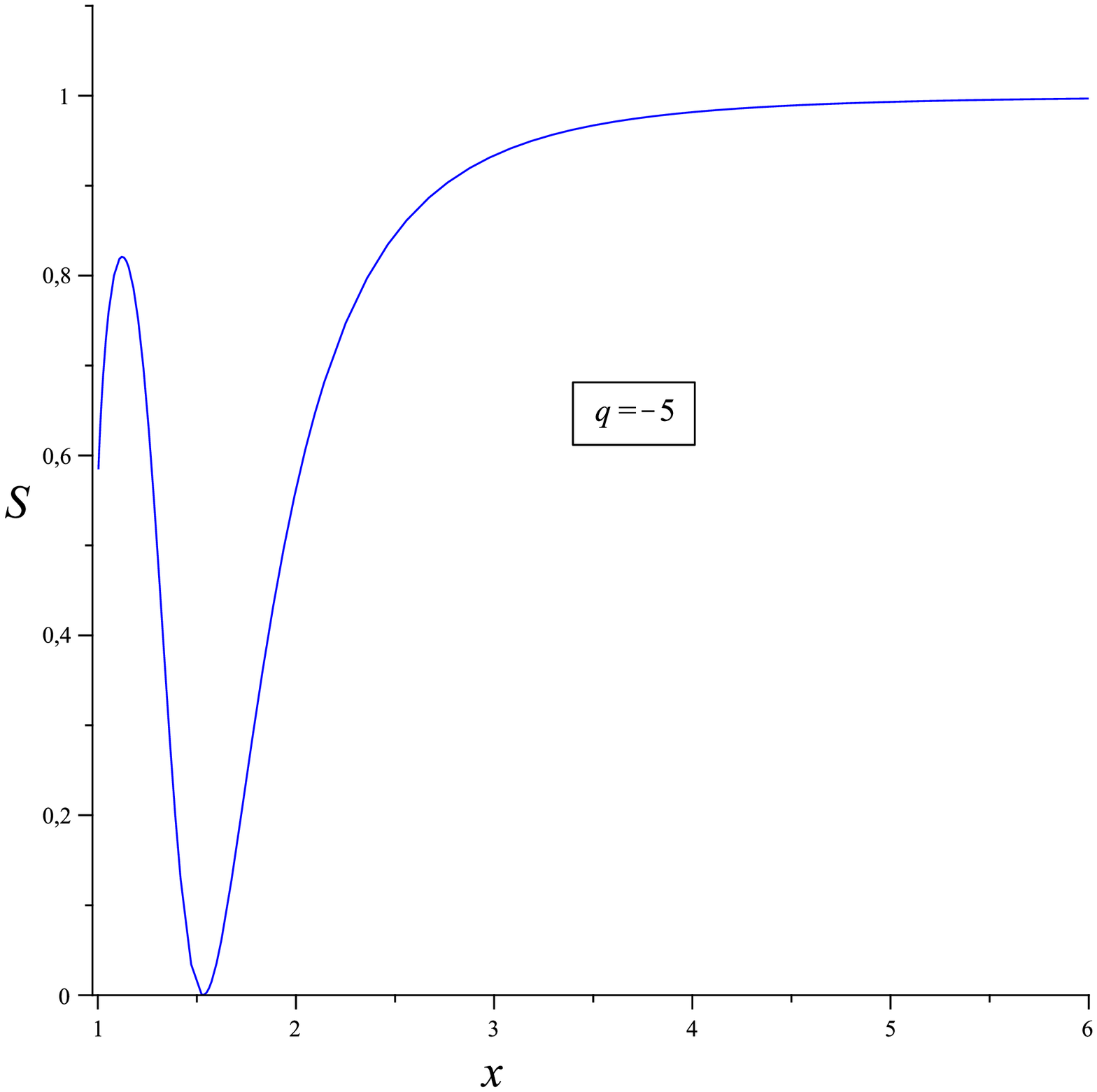}\\[.4cm]
\quad\mbox{(a)}\quad &\quad \mbox{(b)}
\end{array}$\\
\end{center}
\caption{The speciality index ${\mathcal S}$ is plotted as a function of $x$ for $y=0$ and (a) $q=5$ and (b) $q=-5$.
The asymptotic behavior ${\mathcal S}\to 1$ as $x \to\infty$ shows that in such a limit the metric becomes algebraically special, consistently with the \lq\lq peeling theorem" on the asymptotic decay of the mass multipoles of the source.
}
\label{fig:spect}
\end{figure}

The behavior of the curvature invariants $I$ and $J$ as functions of $x$ on the symmetry plane $y=0$ is shown in Figs.~\ref{fig:Iqvari} and \ref{fig:Jqvari} for different values of the quadrupole parameter.
We see how the presence of the quadrupole moment causes $J$ to change sign close to $x=1$, in contrast to the Schwarzschild case ($q=0$).
Correspondingly, the speciality index ${\mathcal S}$ exhibits an oscillating behavior as shown in Fig.~\ref{fig:spect}.

Concerning the physical meaning,  the invariant $I$, which is proportional to the Kretschmann invariant, is most reasonably  related to the strength of the gravitational field under the combined effects of the mass $M$ and its quadrupole $q$.
The physical significance of $J$ is still questionable; however, its oscillations, which riverberate in the behavior of $\mathcal{S}$ and are generated by the presence of the quadrupole parameter $q$, seem to infer the existence of energy bound states in a scattering problem. The explicit expressions of $I$ and $J$ in the case of the Erez-Rosen solution are rather long and not very enlightening so we shall illustrate their properties from the inspection of their plots.


The presence of the quadrupole moment affects the dynamics in a substantial way. The geodesic equations for massive test particles read
\begin{eqnarray}\fl\quad
\label{geoeqns}
\dot t&=&\frac{E}{f}\,, \qquad
\dot\phi=\frac{fL}{\sigma^2X^2Y^2}\,, \nonumber\\
\fl\quad
\ddot y&=&-\frac12\frac{Y^2}{X^2}\left[\frac{f_y}{f}-2\gamma_y+\frac{2y}{X^2+Y^2}\right]{\dot x}^2
+\left[\frac{f_x}{f}-2\gamma_x-\frac{2x}{X^2+Y^2}\right]{\dot x}{\dot y}\nonumber\\
\fl\quad
&&+\frac12\left[\frac{f_y}{f}-2\gamma_y-\frac{2y}{X^2+Y^2}\frac{X^2}{Y^2}\right]{\dot y}^2\nonumber\\
\fl\quad
&&-\frac12\frac{e^{-2\gamma}}{f\sigma^4X^2Y^2(X^2+Y^2)}
[Y^2(f^2L^2+E^2\sigma^2X^2Y^2)f_y+2L^2f^3y]\,, \nonumber\\
\fl\quad
{\dot x}^2&=&-\frac{X^2}{Y^2}{\dot y}^2+\frac{e^{-2\gamma}X^2}{\sigma^2(X^2+Y^2)}\left[E^2-\mu^2f-\frac{f^2L^2}{\sigma^2X^2Y^2}\right]\,,
\end{eqnarray}
where the Killing symmetries and the normalization condition on the tangent vector ${\dot x}^\alpha$ have been used.
Here $E$ and $L$ are the constant energy and angular momentum of the test particle respectively, $\mu$ is the particle mass and a dot denotes differentiation with respect to an affine parameter; furthermore, the notation
\beq
X=\sqrt{x^2-1}\,, \quad Y=\sqrt{1-y^2}\
\eeq
has been introduced and partial derivatives of a generic function $P(x,y)$ with respect to $x$ and $y$ are indicated by 
$P_x$ and $P_y$, respectively.

If $y=0$ and $\dot y=0$ initially, Eqs. (\ref{geoeqns}) ensure that the motion will be confined on the symmetry plane, since $f_y$ and $\gamma_y$ both vanish at $y=0$, so that $\ddot y=0$ as well.
Eqs. (\ref{geoeqns}) thus reduce to
\beq\fl\quad
\label{geoeqnsequat}
\dot t=\frac{E}{f}\,,\qquad
\dot \phi=\frac{fL}{\sigma^2X^2}\,, \qquad
{\dot x}^2=\frac{e^{-2\gamma}X^2}{\sigma^2(1+X^2)}\left[E^2-\mu^2f-\frac{f^2L^2}{\sigma^2X^2}\right]\,,
\eeq
where the metric functions are meant to be evaluated on $y=0$.
Time-like geodesics on the symmetry plane have been investigated in Refs. \cite{quev90,armenti,def}.
In the next section we shall analyze in more detail the null geodesics (i.e., $\mu=0$) on the symmetry plane $y=0$.

\subsection{Light scattering on the symmetry plane}

\begin{figure}
\begin{center}
\includegraphics[scale=0.35]{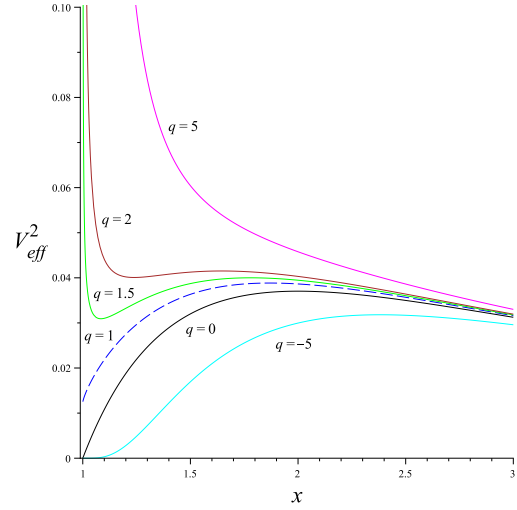}
\end{center}
\caption{The behavior of the square of the effective potential (\ref{eqVx}) as a function of $x$ is shown for different values of the quadrupole parameter.
Asymptotically as $x\to\infty$ all curves tend to zero.
The dashed curve corresponds to $q=1$ and denotes a separatrix discriminating the behavior of $V_{\rm (eff)}^2$ for $x\to1$.
In fact, we have $\lim_{x\to1^+} V_{\rm (eff)}^2=0$ if $q<1$ and $\lim_{x\to1^+} V_{\rm (eff)}^2=\infty$ if $q>1$; when $q=1$ it tends to the constant value $1/(4e^3)$. Notice that for values of $q\gtrsim1$ there exist bounded photon orbits nearby the singularity.
}
\label{fig:Veff}
\end{figure}

\begin{figure}
\begin{center}
$\begin{array}{cc}
\includegraphics[scale=0.3]{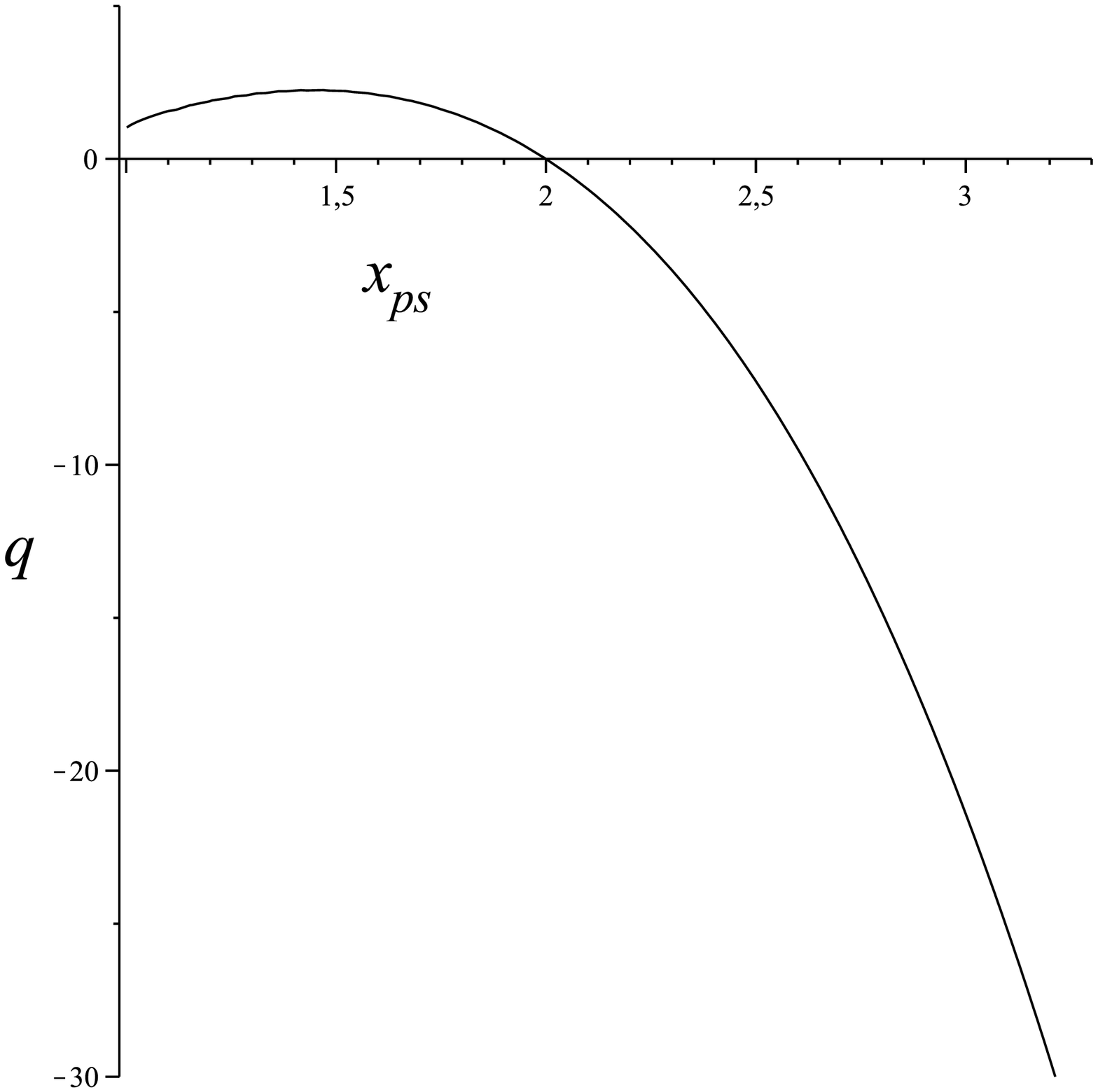}&\quad
\includegraphics[scale=0.3]{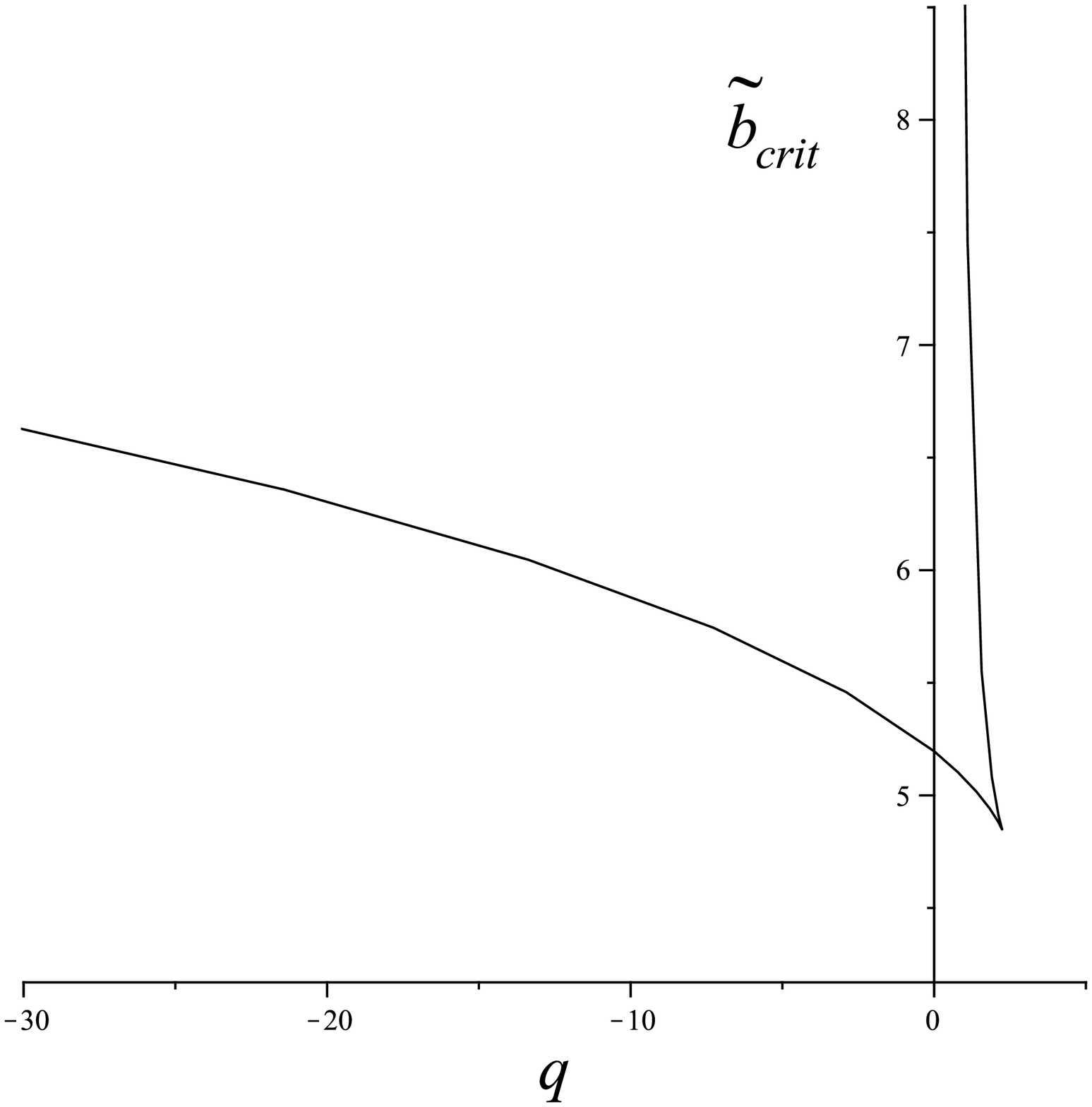}\\[.4cm]
\quad\mbox{(a)}\quad &\quad \mbox{(b)}
\end{array}$\\
\end{center}
\caption{
The location of the photon sphere $x=x_{\rm ps}$, i.e., the extremal points of the effective potential, and the corresponding critical impact parameter as functions of $q$ are shown in panel (a) and (b), respectively.
The curve ${\tilde b}_{\rm crit}$ versus $q$ has two branches in the range $1<q\lesssim 2.254$, where two different photon sphere for a given value of $q$ are present.
}
\label{fig:bandxcrit}
\end{figure}

Consider a photon moving on the symmetry plane.
Its trajectory is described by the equation
\beq
\label{normalizx}
\left(\frac{\rmd x}{\rmd\phi}\right)^2
=e^{-2q(qG_2+G_1+F)}(x+1)^4\left[\frac{M^2}{b^2}-\frac{x-1}{(x+1)^3}\,e^{2qF}\right]\,,
\eeq
where
\begin{eqnarray}
F&=&-\frac14(3x^2-1)\ln\left(\frac{x-1}{x+1}\right)-\frac32x\,, \nonumber\\
G_1&=&\ln\left(\frac{x^2-1}{x^2}\right)-\frac32x\ln\left(\frac{x-1}{x+1}\right)-3\,, \nonumber\\
G_2&=&\frac12\ln\left(\frac{x^2-1}{x^2}\right)+\left[\frac38(x^2-1)\ln\left(\frac{x-1}{x+1}\right)\right]^2\nonumber\\
&&+\frac{3}{16}x(3x^2-5)\ln\left(\frac{x-1}{x+1}\right)+\frac3{16}(3x^2-4)\,,
\end{eqnarray}
and $b=L/E$ denotes the impact parameter.
From Eq. (\ref{normalizx}) we deduce that the motion is governed by the effective potential
\beq
\label{eqVx}
V_{\rm (eff)}=\frac{e^{qF}}{x+1}\sqrt{\frac{x-1}{x+1}}\,,
\eeq
whose behavior as a function of $x$ is shown in Fig.~\ref{fig:Veff} for different values of $q$.
Eq. (\ref{normalizx}) rewrites as
\beq
\label{normalizx2}
\left(\frac{\rmd x}{\rmd\phi}\right)^2
=e^{-2q(qG_2+G_1+F)}(x+1)^4\left[\frac{M^2}{b^2}-V_{\rm (eff)}^2\right]\,,
\eeq
so that for fixed values of $\tilde b=b/M$ the physical motion is confined to the values of $x$ such that $\tilde b^{-2}\ge V^2_{\rm (eff)}$.
Solving the equation $\tilde b^{-2}= V^2_{\rm (eff)}$ for $x$ gives the turning points associated with the selected orbit.
The special case of spatially circular motion is obtained by setting to zero the first derivative of the effective potential with respect to $x$, i.e.,
\beq
\label{photosphere}
\frac{\rmd V_{\rm (eff)}}{\rmd x}=0\qquad \to \qquad
q(x^2-1)F'-x+2=0\,,
\eeq
where the apex means derivative with respect to $x$.
Numerical study of this equation shows that it admits either a single root or two roots or even no roots depending on the value of the quadrupole parameter $q$ (see Figs.~\ref{fig:Veff} and~\ref{fig:bandxcrit} (a)).
Correspondingly, $V_{\rm (eff)}$ exhibits a single maximum for $q\le1$, both a maximum and a minimum in the range $1<q\lesssim 2.254$, and is monotonically decreasing with $x$ for $q\gtrsim2.254$.
These extremal points mark spatially circular photon orbits, or photon spheres in the terminology of Virbhadra and Ellis \cite{ellis2}, and will be denoted by $x=x_{\rm ps}$.
They have shown how the presence of a photon sphere affects the lensing properties of the background metric and have also provided a classification of naked singularities depending on whether or not the naked singularity is covered by a photon sphere: weakly naked singularities (WNS) are those contained within at least one photon sphere, whereas strongly naked singularities (SNS) are those not contained within any photon sphere.\footnote{
This classification should be confronted with the one given in Ref. \cite{defussbra}.
}
According to this classification, in the case of the Erez-Rosen solution, the naked singularity would be strong for $q\gtrsim2.254$ and weak otherwise.

The value of the impact parameter corresponding to a spatially circular photon orbit is denoted by ${\tilde b}_{\rm crit}$, whose dependence on $q$ is shown in Fig.~\ref{fig:bandxcrit} (b). Remarkably, the presence of a mass quadrupole allows
 for both stable and unstable photon spheres.
In the Schwarzschild case ($q=0$), the above family of orbits, described by Eq. (\ref{photosphere}), merge into an unstable one at $x=2$, i.e., $r=3M$, with a critical impact parameter ${\tilde b}_{\rm crit}=3\sqrt{3}\approx 5.196$.

\begin{figure}
\begin{center}
$\begin{array}{cc}
\includegraphics[scale=0.35]{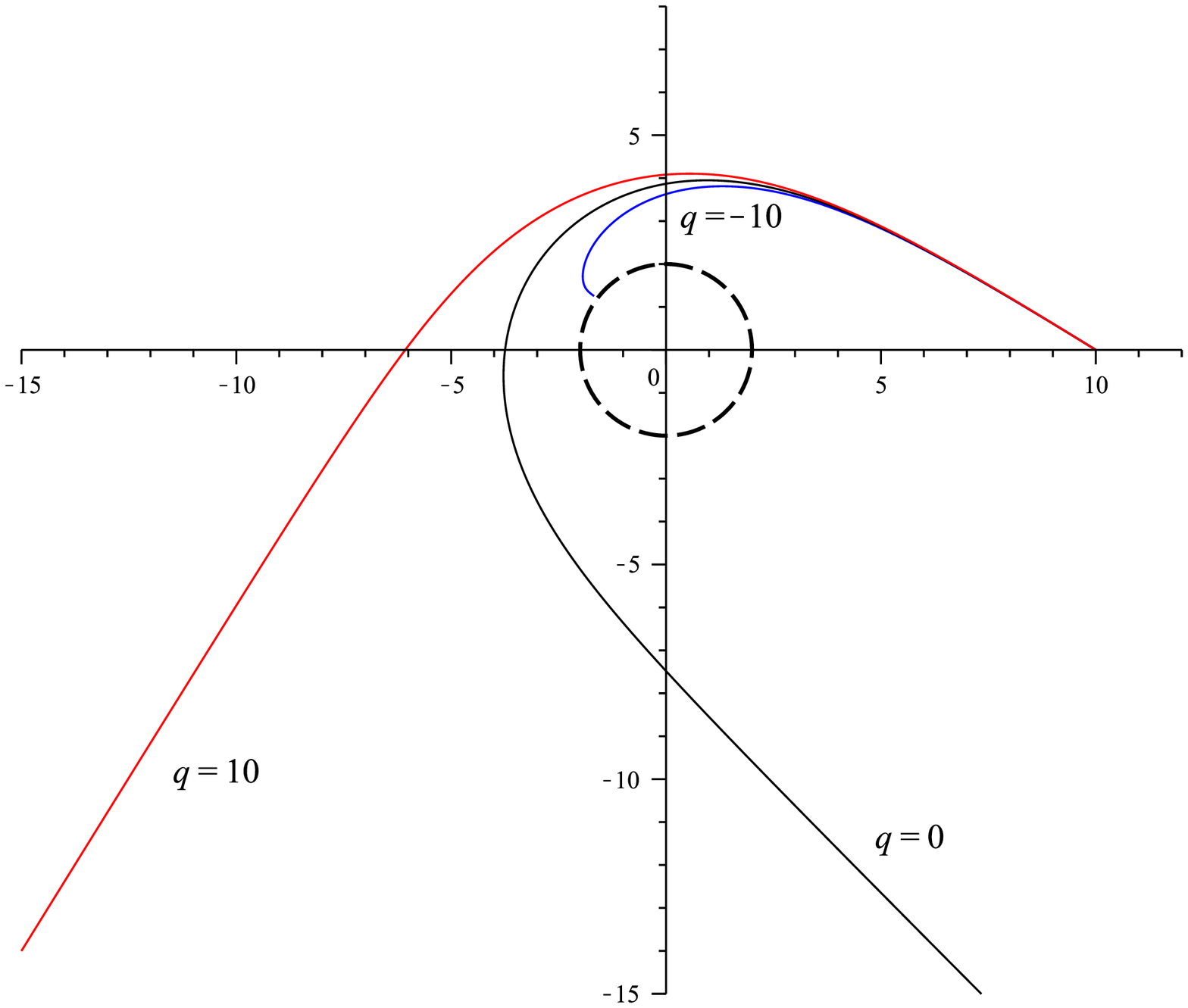}&\quad
\includegraphics[scale=0.35]{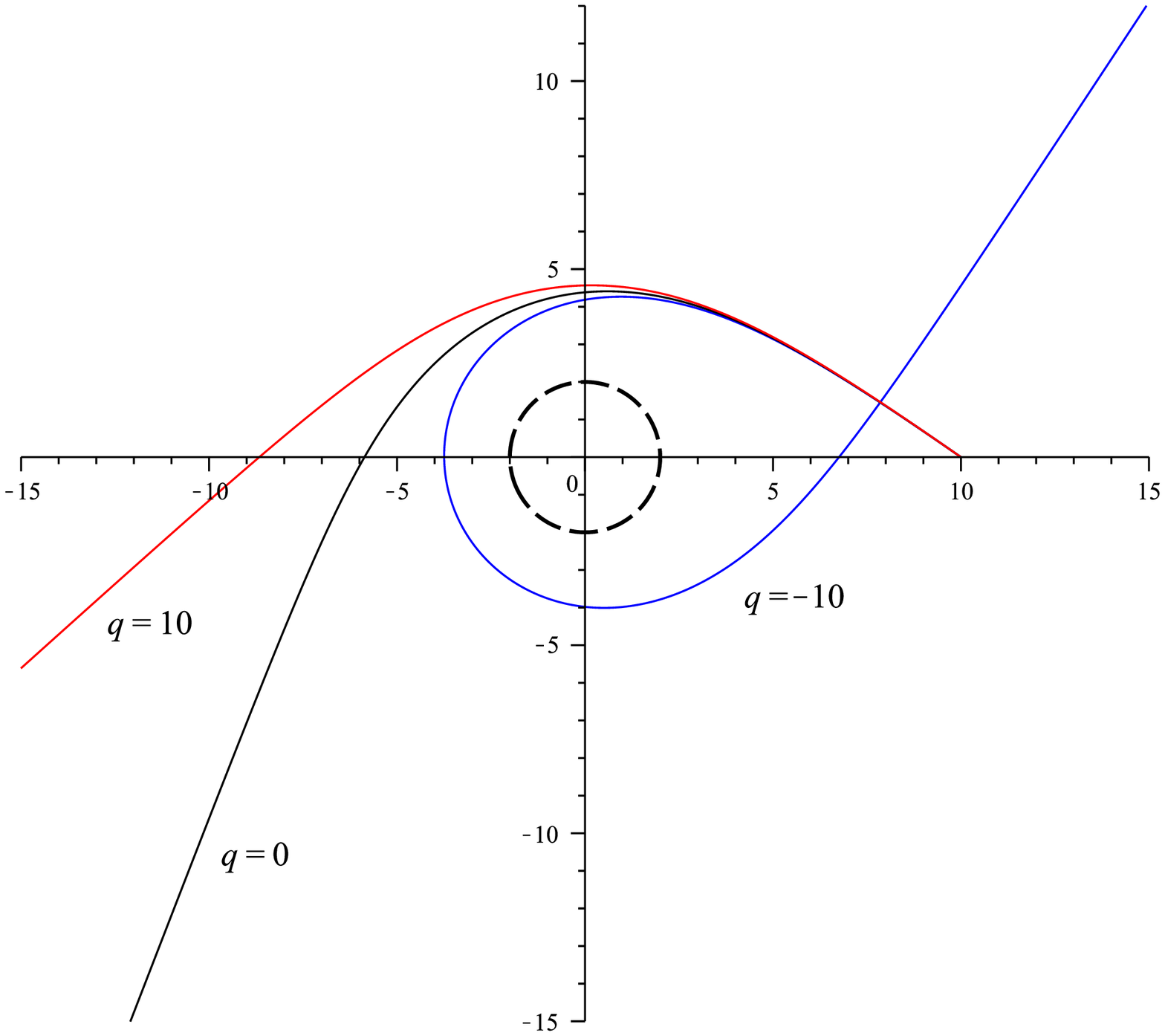}\\[.4cm]
\quad\mbox{(a)}\quad &\quad \mbox{(b)}
\end{array}$\\
\end{center}
\caption{Photon orbits on the symmetry plane for (a) $b/M=5.4$ and (b) $b/M=5.9$ and different values of $q=[-10,0,10]$.
Cartesian-like coordinates $[(r/M)\cos\phi,(r/M)\sin\phi]$ have been used.
}
\label{fig:orbite}
\end{figure}

The critical value of the impact parameter plays a central role in characterizing the allowed orbits of different kinds.
In fact, a photon with an impact parameter $\tilde b$ will be scattered by the source on unbound orbits or be captured or eventually be accomodated on spatially circular orbits either stable or unstable, according to the relative values of $\tilde b$ and $\tilde b_{\rm crit}$.
Examples of numerical integration of orbits are shown in Fig.~\ref{fig:orbite}.

\begin{figure}
\begin{center}
\includegraphics[scale=0.35]{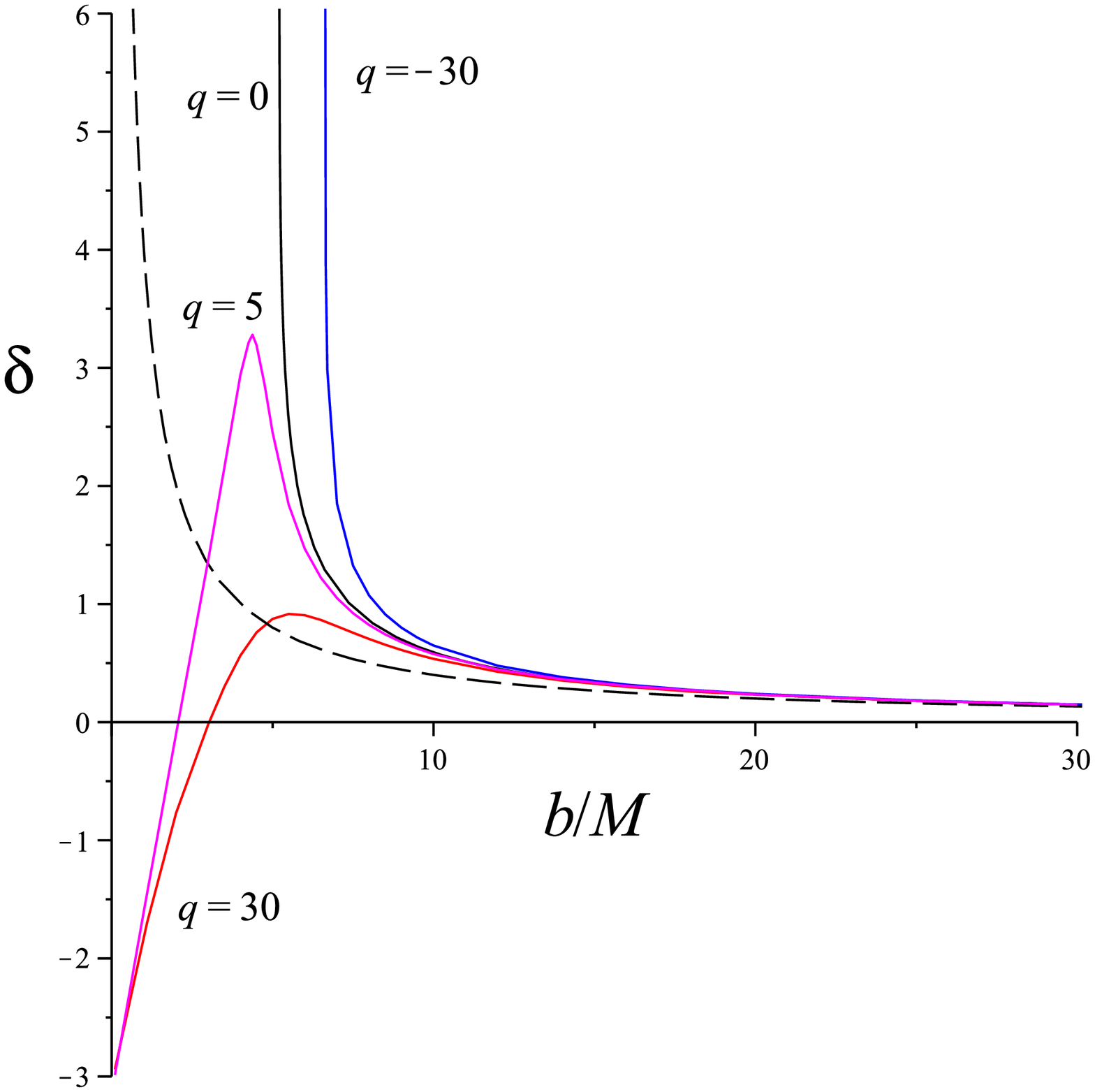}
\end{center}
\caption{Deflection angle as a function of the impact parameter for different values of $q=[-30,0,5,30]$.
The black dashed line correspond to the standard result $\delta\sim4M/b$ in the limit of weak deflection in a Schwarzschild field.
Deviations from the asymptotic behavior depend on the quadrupole parameter in a significant way at about $b/M\sim10$, leading to a bifurcation as approaching smaller values of $b/M$.
}
\label{fig:delta}
\end{figure}

In the case of open orbits one can define the deflection angle as the angle between the asymptotic incoming and outgoing trajectories, i.e.,
\beq
\label{deflectionangle}
\delta=2\int_{x_{\rm min}}^\infty \left(\frac{\rmd\phi}{\rmd x}\right)\,\rmd x -\pi\,,
\eeq
by using Eq. (\ref{normalizx}).
The behavior of $\delta$ as a function of $\tilde b$ is shown in Fig.~\ref{fig:delta} for different values of $q$.
For negative values of $q$ the deflection angle strictly increases with the decrease of the impact parameter and becomes unboundedly large as the impact parameter approaches its corresponding value on the photon sphere, i.e., for $b\to b_{\rm crit}$.
For positive values of $q\gtrsim2.254$, instead, the singularity is no longer covered by any photon sphere, implying that the deflection angle for these cases is never unboundedly large. It increases up to its maximum for decreasing values of the impact parameter, then decreases for $b$ further decreasing and changes its sign until it finally reaches its minimum value as the impact parameter approaches the limiting value $b\to0$ (for $x\to1$).
This is a characteristic feature of SNS, as pointed out in Ref. \cite{ellis2}.
It is worth noticing here such a peculiar behavior of $\delta$ for positive values of $q$.
As expected, $\delta$ increases as $b$ decreases, since the incoming photon feels a stronger gravitational field as it approaches the gravity source. The latter, however, being prolate, has its mass mainly concentrated along the axis $y=\pm1$, therefore a photon moving in the symmetry plane $y=0$, once is sufficiently close to the gravity source,  will feel a decreasing gravitational field since  most of the gravitational action is being neutralized between the North and the South mass distribution relative to the plane $y=0$. Yet, a better understanding of the optical properties of the given metric
in view of its application to astrometric problems, is assured by analyzing its lensing effects.


\subsection{Gravitational lensing in the weak deflection limit}

\begin{figure}[h]
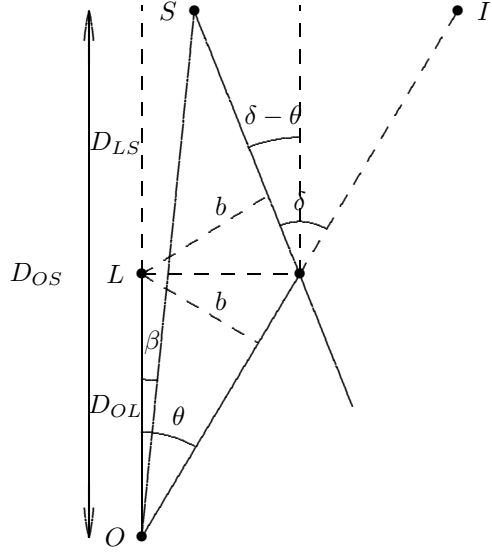

$$ \vbox{
\beginpicture

  \setcoordinatesystem units <0.7cm,0.7cm> point at 0 0
\putrule from 0 0 to  0 10
\putrule from 1 0 to  1 5

\plot  1 0    2 10    /
\plot  1 0    4 5    /
\plot  4 5    2 10     /
\plot  4 5    5 2.5     /

\setdashes
\putrule from 1 5 to  4 5
\putrule from 1 5 to  1 10.1
\putrule from 4 5 to  4 10.1

\plot  4 5    7 10     /
\plot  1 5    3.2 3.7     /
\plot  1 5    3.4 6.5     /

\put {\mathput{\bullet}}              at 1 5
\put {\mathput{\bullet}}              at 2 10 
\put {\mathput{\bullet}}              at 7 10 
\put {\mathput{\bullet}}              at 1 0  
\put {\mathput{\bullet}}              at 4 5

\put {\mathput{D_{OS}}}              at -1 5
\put {\mathput{D_{LS}}}              at 0.5 7.5
\put {\mathput{D_{OL}}}              at 0.5 2.5

\put {\mathput{L}}               at 0.5 5
\put {\mathput{S}}               at 1.5 10
\put {\mathput{I}}               at 7.5 10
\put {\mathput{O}}               at 0.5 0
\put {\mathput{\delta }}               at 4 6.4
\put {\mathput{\theta }}               at 1.7 2.3
\put {\mathput{\beta }}               at 1.2 3.7
\put {\mathput{b}}               at 2.5 4.5
\put {\mathput{b}}               at 2.5 6.3
\put {\mathput{\delta-\theta }}               at 3.5 8

\setsolid
\circulararc 22 degrees from 4 6 center at 4 5
\circulararc -30 degrees from 4 6 center at 4 5
\circulararc -30 degrees from 1 2 center at 1 0
\circulararc -6 degrees from 1 3 center at 1 0
\circulararc 22 degrees from 4 7.6 center at 4 5

\arrow <.4cm> [.2,.3]    from  0 9  to 0 10
\arrow <.4cm> [.2,.3]    from  0 1  to 0 0

\endpicture}$$

\caption{Schematic diagram of the lens geometry.
$O$, $L$, $S$ and $I$ denote the positions of the observer, lens, source and primary image, respectively.
The line joining the observer and the lens is the reference optical axis, from which the angular separations of the source ($\beta$) and the image ($\theta$) are measured.
}
\label{fig:lens}
\end{figure}

Let us study the gravitational lensing of a distant star (hereafter the source $S$) with respect to an observer $O$ by a quasi-spherical body (the lens $L$) described by the Erez-Rosen solution.
Referring to the diagram shown in Fig.~\ref{fig:lens}, the geometry of the system implies
\beq
D_{OL}\tan\theta=D_{OS}\tan\beta+D_{LS}\tan(\delta-\theta)\,,
\eeq
where $D_{LS}$, $D_{OS}$ and $D_{OL}=D_{OS}-D_{LS}$ are the lens-source, the observer-source and the observer-lens distances, respectively.
The equation of the gravitational lens thus becomes (see, e.g., Ref. \cite{ellis} and references therein)
\beq
\label{lenseq}
\tan\beta=\tan\theta-\frac{D_{LS}}{D_{OS}}[\tan\theta+\tan(\delta-\theta)]\,.
\eeq
The line joining the observer $O$ and the lens $L$ is taken as the reference optical axis, from which the angular positions of the source and the image, denoted by $\beta$ and $\theta$ respectively, are measured.
The deflection angle $\delta$ is given by Eq. (\ref{deflectionangle}).
From the lens geometry, the impact parameter $b$ of the light ray is related to the angle $\theta$ by
\beq
\label{bdith}
b=D_{OL}\sin\theta\,,
\eeq
so yielding the dependence of the deflection angle on the variable $\theta$.

The locations of the images are given by those values of $\theta$ satisfying the lens equation (\ref{lenseq}) for fixed values of the source position $\beta$.
The signed magnification of an image is defined by
\beq
\label{mudef}
\mu=\left(\frac{\sin\beta}{\sin\theta}\frac{\rmd\beta}{\rmd\theta}\right)^{-1}\,.
\eeq

In the weak deflection limit, the position of the images as well as their magnification can be determined by calculating the solutions of an appropriate series expansion of the general lens equation (\ref{lenseq}), order by order.
According to Ref. \cite{keeton}, it is useful to re-scale both angles $\beta$ and $\theta$ by the Einstein radius
\beq
\theta_E=\sqrt{\frac{4MD_{LS}}{D_{OL}D_{OS}}}\,,
\eeq
and to introduce the expansion parameter $\xi=\theta_ED_{OS}/(4D_{LS})$.
Therefore, the new variables in the lens equation are $\tilde\beta=\beta/\theta_E$ and $\tilde\theta=\theta/\theta_E$,  and the solution can be written as a series expansion of the form
\beq
\label{thexp}
\tilde\theta=\tilde\theta_0+\xi\tilde\theta_1+\xi^2\tilde\theta_2+O(\xi^3)\,,
\eeq
where $\tilde\theta_0$ is the well known image position in the lowest-order weak deflection limit, and the coefficients $\tilde\theta_1$ and $\tilde\theta_2$ give the correction terms which remain to be determined by solving the first and second order approximate lens equation respectively, for a fixed value of the source position $\tilde\beta$.

The weak field expansion of the deflection angle (\ref{deflectionangle}) is
\beq
\delta=\frac{4M}{b}+\frac{15\pi}{4}\frac{M^2}{b^2}+\left(\frac{128}{3}-\frac{8}{15}q\right)\frac{M^3}{b^3}+O\left(\frac{M^4}{b^4}\right)\,,
\eeq
where $b$ is given by Eq. (\ref{bdith}).
Note that this approximate formula is valid in the regime of large values of $b/M$, i.e., in the rightmost region of Fig.~\ref{fig:delta} where the curves approach the asymptotic weak deflection limit (see Ref. \cite{matzner}).
This relation reduces to the well known light deflection results in the Schwarzschild case (see, e.g., Ref. \cite{ellis}).
Introducing the parameter $\xi$ and making the expansion (\ref{thexp}) we find
\begin{eqnarray}\fl\quad
\label{deltaexp}
\delta&=&\frac{4\xi}{\tilde\theta_0}\left\{
1+\left(\frac{15\pi}{16}-\tilde\theta_1\right)\frac{\xi}{\tilde\theta_0}\right.\nonumber\\
\fl\quad
&&\left.+\left[\tilde\theta_1^2-\frac{15\pi}{8}\tilde\theta_1
+\frac{8}{3}\frac{D_{LS}^2}{D_{OS}^2}\tilde\theta_0^4-\tilde\theta_0\tilde\theta_2
+\frac{32}{3}-\frac{2}{15}q\right]\frac{\xi^2}{\tilde\theta_0^2}\right\}+O(\xi^4)\,.
\end{eqnarray}
Substituting then in the lens equation, Taylor expanding in $\xi$ and solving order by order finally give
\begin{eqnarray}\fl\quad
\tilde\theta_0&=&\frac{1}{2}(\tilde\beta\pm\sqrt{\tilde\beta^2+4})\equiv\tilde\theta_0^{\pm}\,,\qquad
\tilde\theta_1=\frac{15\pi}{16(1+\tilde\theta_0^2)}\,,\nonumber\\
\fl\quad
\tilde\theta_2&=&\frac{1}{\tilde\theta_0(1+\tilde\theta_0^2)}\left[
\frac{8}{3}\frac{D_{LS}^2}{D_{OS}^2}(\tilde\theta_0^4+6\tilde\theta_0^2-2)
-16\tilde\theta_0^2\frac{D_{LS}}{D_{OS}}
+\frac{16\tilde\theta_0^4}{(1+\tilde\theta_0^2)^2}\right.\nonumber\\
\fl\quad
&&\left.+\frac{1+2\tilde\theta_0^2}{(1+\tilde\theta_0^2)^2}\left(\frac{225\pi^2}{256}-16\right)
-\frac{2}{15}q
\right]\,.
\end{eqnarray}
The $\pm$ signs correspond to the primary $(+)$ and secondary $(-)$ images forming on either side of the source, with the former one outside the Einstein ring and the latter inside.
Note that $\tilde\beta$ is assumed to be positive in both cases, whereas $\tilde\theta$ is positive/negative for positive/negative-parity images.
Therefore, the convention used here is different from that of Ref. \cite{keeton}.
The expansion of the signed magnification (\ref{mudef}) of each individual image is
\beq
\mu=\mu_0+\xi\mu_1+\xi^2\mu_2+O(\xi^3)\,,
\eeq
with
\begin{eqnarray}\fl\quad
\mu_0&=&\frac{\tilde\theta_0^4}{\tilde\theta_0^4-1}\,,\qquad
\mu_1=-\frac{15\pi\tilde\theta_0^3}{16(1+\tilde\theta_0^2)^3}\,,\nonumber\\
\fl\quad
\mu_2&=&\frac{\tilde\theta_0^4}{(1+\tilde\theta_0^2)^3(\tilde\theta_0^2-1)}\left[
\frac{8}{3}\frac{D_{LS}^2}{D_{OS}^2}(\tilde\theta_0^4+16\tilde\theta_0^2+1)
-32\left(\frac{D_{LS}}{D_{OS}}+1\right)\right.\nonumber\\
\fl\quad
&&\left.+\frac{675\pi^2\tilde\theta_0^2}{128(1+\tilde\theta_0^2)^2}
+\frac{4}{15}q
\right]\,.
\end{eqnarray}

When the two weak field images are too close together to be resolved, the main observables become the total magnification and magnification-weighted centroid position.
The total magnification turns out to be given by
\begin{eqnarray}\fl\quad
\mu_{\rm tot}&=&|\mu^+|+|\mu^-|\nonumber\\
\fl\quad
&=&\frac{\tilde\beta^2+2}{\tilde\beta\sqrt{\tilde\beta^2+4}}
-\frac{15\pi}{8(\tilde\beta^2+4)^{3/2}}\xi\nonumber\\
\fl\quad
&&+\frac{2}{\tilde\beta(\tilde\beta^2+4)^{3/2}}\left[
\frac{8}{3}\frac{D_{LS}^2}{D_{OS}^2}(18+\tilde\beta^2)
-32\left(\frac{D_{LS}}{D_{OS}}+1\right)\right.\nonumber\\
\fl\quad
&&\left.+\frac{675\pi^2}{128(\tilde\beta^2+4)}
+\frac{4}{15}q
\right]\xi^2+O(\xi^3)\,,
\end{eqnarray}
taking into account that the image parities give for the absolute magnifications $|\mu^+|=\mu^+$ and $|\mu^-|=-\mu^-$.
The magnification-weighted centroid position is instead given by
\begin{eqnarray}\fl\quad
\Theta_{\rm cent}&=&\frac{\theta^+|\mu^+|+\theta^-|\mu^-|}{|\mu^+|+|\mu^-|}\nonumber\\
\fl\quad
&=&\frac{\tilde\beta(\tilde\beta^2+3)}{\tilde\beta^2+2}
-\frac{15\pi(\tilde\beta^2+1)}{8(\tilde\beta^2+2)^2}\xi\nonumber\\
\fl\quad
&&+\frac{\tilde\beta}{(\tilde\beta^2+2)^2}\left[
\frac{8}{3}\frac{D_{LS}^2}{D_{OS}^2}(\tilde\beta^4+9\tilde\beta^2-2)
-16\left(\frac{D_{LS}}{D_{OS}}\tilde\beta^2-2\right)\right.\nonumber\\
\fl\quad
&&\left.-\frac{225\pi^2}{128(\tilde\beta^2+2)}
-\frac{4}{15}q
\right]\xi^2+O(\xi^3)\,.
\end{eqnarray}

We have developed up to now a semi-analytical study of the photon orbits in the general case of the exact Erez-Rosen metric, where fully analytic calculations are complicated by the quadrupolar structure of the central body. We have then considered the corresponding weak field limit, in which Taylor expanded results have been obtained for both light deflection and lensing observables.

For the sake of completeness, let us discuss how to exploit the Erez-Rosen solution also in the case one wants to determine the photo-centric position of a star. 
In order to discuss fully analytical results (like the photon orbit, i.e., the location of the star in a sky map), a fact which is necessary to perform relativistic astrometric models especially in view of dedicated missions like Gaia, we pass now to reconsider the above discussion in the weak field case, having in mind the characteristic features induced by the strong field regime of the Erez-Rosen solution.

\subsection{The weak field and small quadrupole approximation in harmonic coordinates}

The weak field limit and small quadrupole approximation of the Erez-Rosen solution expressed in harmonic coordinates require some care.
The exterior field of a slowly rotating slightly deformed mass is described by the Hartle-Thorne metric \cite{ht67}, which is an approximate solution of the Einstein's field equation, accurate to second order in the rotation parameter and to first order in the mass quadrupole moment.
The Hartle-Thorne solution is a generalization of the Lense-Thirring spacetime \cite{lt18,hmt}; it
has been obtained by a detailed consideration of the interior structure of the astronomical object together with a proper matching of interior and exterior solutions at the boundary.

First of all, we show that, once linearized with respect to the quadrupole parameter, the Erez-Rosen metric can be transformed into the Hartle-Thorne one, in the absence of rotation, since both of them belong to a more general class of solutions which can be obtained from the Erez-Rosen solution by the Zipoy-Voorhees transformation (see also Refs. \cite{mt91,bglq}). Details are in 
\ref{ERtoHT}.
The standard form of the non-rotating Hartle-Thorne solution in Schwarzschild-like coordinates $(t,R,\Theta,\phi)$ is given by
\begin{eqnarray}\fl
\label{HTmet}
\rmd s^2&=&-\left(1-\frac{2{\mathcal M}}{R}\right)\left[
1+2k_1P_2(\cos\Theta)\right]\rmd t^2
+\left(1-\frac{2{\mathcal M}}{R}\right)^{-1}\left[
1-2k_1P_2(\cos\Theta)\right]\rmd R^2\nonumber\\
\fl
&&+R^2(\rmd\Theta^2+\sin^2\Theta\rmd\phi^2)[1-2k_2P_2(\cos\Theta)]\,,
\end{eqnarray}
where 
\begin{eqnarray}
k_1&=&-\frac58\frac{{\mathcal Q}}{{\mathcal M}^3}Q_2^2\left(\frac{R}{{\mathcal M}}-1\right)\ , \nonumber\\
k_2&=&k_1-\frac54\frac{{\mathcal Q}}{{\mathcal M}^2R}\left(1-\frac{2{\mathcal M}}{R}\right)^{-1/2}Q_2^1\left(\frac{R}{\mathcal M}-1\right)\ . 
\end{eqnarray}
Here $Q_l^m$ are the associated Legendre functions of the second kind and the constants ${\mathcal M}$ and ${\mathcal Q}$ are the total mass and quadrupole moment of the body, respectively.
\footnote{
The mass  ${\mathcal M}$ and quadrupole moment ${\mathcal Q}$ entering the Hartle-Thorne solution are related to the corresponding parameters $M$ and $q$ of the Erez-Rosen solution by the relations ${\mathcal M}=M(1-q)$ and ${\mathcal Q}=(4/5)M^3q$, as indicated below in \ref{ERtoHT}.
}

Next, we conveniently express the Hartle-Thorne solution (i.e., our approximated Erez-Rosen solution) in harmonic coordinates.
To this end, let us start by expressing the mass quadrupole moment ${\mathcal Q}=J_2{\mathcal M}d^2$ in the Hartle-Thorne metric (\ref{HTmet}) in terms of a new dimensionless quadrupole parameter $J_2$ and the equatorial radius $d$ of the body.
Restoring then physical units in the Hartle-Thorne solution by replacing  ${\mathcal M}\to G{\mathcal M}/c^2$
and considering  the expansion of the metric up to terms of the order  $1/c^2$ included  
leads to
\begin{eqnarray}\fl\quad
\label{ERapprox}
\rmd s^2 &=& -\left[1-\frac{2G{\mathcal M}}{c^2R}-2J_2\frac{G{\mathcal M}d^2}{c^2R^3}P_2(\cos\Theta)\right]\rmd t^2\nonumber\\
\fl\quad
&&+\left[1+\frac{2G{\mathcal M}}{c^2R}+2J_2\frac{G{\mathcal M}d^2}{c^2R^3}P_2(\cos\Theta)\right]\rmd R^2\nonumber\\
\fl\quad
&&+R^2\left[1+2J_2\frac{G{\mathcal M}d^2}{c^2R^3}P_2(\cos\Theta)\right](\rmd \Theta^2+\sin^2 \Theta \rmd \phi^2)\,.
\end{eqnarray}
Harmonic coordinates $x_h^\alpha$ can then be obtained  by the transformation $R=r_{h}+G{\mathcal M}/c^2$, leaving the remaining coordinates unchanged.
In doing so we obtain the following weak field limit and small quadrupole approximation of the Erez-Rosen metric in harmonic coordinates
\beq\fl\qquad\label{PNmetric}
\rmd s^2 = -(1-\epsilon^2h)(\rmd x^0)^2 +(1+\epsilon^2h)[\rmd r_{h}^2 +r_{h}^2(\rmd \theta_{h}^2+\sin^2 \theta_{h} \rmd \phi_{h}^2)]\,,
\eeq
where $x^0=ct_{h}$, $\epsilon=1/c$ and
\beq
h=\frac{2G{\mathcal M}}{r_{h}}\left(1+J_2\frac{d^2}{r_{h}^2}P_2(\cos\theta_{h})\right)\,.
\eeq
Noticeably, the form (\ref{PNmetric}) of the metric also satisfies the well known \lq\lq conformal isotropy condition"
\beq
-g_{00}g_{ij}\rmd x_h^i \rmd x_h^j=\rmd r_{h}^2 +r_{h}^2(\rmd \theta_h^2+\sin^2 \theta_h \rmd \phi_h^2)+O\left( \frac{1}{c^4} \right)\,.
\eeq

The form (\ref{PNmetric}) of the Erez-Rosen metric in the given approximation appears fully equivalent to the post-Newtonian solution which is exploited in relativistic astrometry within the Solar System. 
In this limiting situation, problems like motion, deflection and lensing of a light ray, examined only numerically in the case of the exact Erez-Rosen solution, can be discussed analytically. This is true, for instance, for 
the inverse ray tracing problem  which we are going to discuss in the next section and which is at the basis of the astrometric problem 
associated with space missions like Gaia.

\section{The approximate Erez-Rosen solution and its generalization to quasi-spherical multipolar sources}

Let us start our analysis with general considerations.
Consider a  stationary $N-$body configuration in a background metric described by the line element
\beq
ds^2 =g_{00} (dx^0)^2+2g_{0i} dx^0 dx^i +g_{ij}dx^i dx^j\,.
\eeq
Here the quasi-Cartesian coordinates $x^0=ct$, where $t$ denotes time, $c$ is the speed of light, and $x^a$ ($a=1,2,3$, $x^1=x, x^2=y, x^3=z$) have the dimensions of a length; we have, to the order of $\epsilon^2$
\beq
\begin{array}{clcl}
\label{metric_coeffs}
g_{00}=& -1+\epsilon^2h+O(4)\,, & g^{00}=& -1-\epsilon^2h+O(4)\,, \\
g_{0i}=&O(3)\,, & g^{0i}=&O(3)\,,  \\
g_{ij}=&\delta_{ij}\left(1+\epsilon^2 h\right)+O(4)\,,\quad & g^{ij}=&\delta^{ij}\left(1-\epsilon^2 h\right)+O(4) \,.
\end{array}
\eeq
If one takes into account the mass multipolar structure of the bodies which must be considered of finite size (and within a common \lq\lq equatorial" plane), then
\beq
\label{hdef}
h= \sum_{a}\frac{2GM_{(a)}}{r_{(a)}} \left[ 1- \sum_{n=2}^{\infty}  J_{(a)\,n}\left(
\frac{d_{(a)} }{r_{(a)}} \right)^n  P_n\left(\frac{z_{(a)}}{r_{(a)}}\right) \right]\,,
\eeq
where $r^i_{(a)}= x^i - x^i_{(a)}$ are the coordinates of the $(a)$-th body with the origin fixed at the center of mass of the whole system, $P_n$ are the Legendre polynomials, $M_{(a)}$ the mass of the $(a)$-th body, $d_{(a)}$ its equatorial radius, as stated, and the coefficients $J_{(a)\,n}$ are the mass multipole moments.\footnote{
Actually modern astrometry requires $\epsilon^3$ order of accuracy. However, for the purpose of this paper it is enough to treat the problem at the $\epsilon^2$ level. 
}

Hereafter, we shall be mainly interested in the quadrupole structure, associated with $J_{(a)\,2}$. Since the following analysis requires a physical interpretation in order to be of some usefulness,  we need to specify a physical observer.
For convenience, we select as fiducial observers  those with $4$-velocity
\beq
\label{obser}
m=\frac1{\sqrt{-g_{00}}}\partial_0\simeq\left(1+\epsilon^2\frac{h}{2}\right)\partial_0\,,
\eeq
and local rest frame spanned by the adapted orthonormal spatial triad
\beq
\label{fram}
e(m)_{\hat a}\simeq\left(1-\epsilon^2\frac{h}{2}\right)\partial_a\,.
\eeq
As required by the object of our investigation, we shall first deduce the trajectory of a light ray perturbed to the order of $\epsilon^2$  by the properties of the source up to its quadrupole.

Tensor components which are first order in $h$ (or, equivalently, to the order of $\epsilon^2$) are lowered/raised with the Minkowski metric.

\subsection{The perturbed photon trajectories}

Let the null geodesic of the photon be described by the tangent vector field $K$, i.e.,
\beq
\label{geo_gen}
K^\alpha\nabla_\alpha K^\beta=0\,, \qquad
K^\alpha K_\alpha=0\,,
\eeq
$\nabla_\alpha$ being the covariant derivative associated with the spacetime metric.
In terms of the observer (\ref{obser}) and (\ref{fram}), $K$ reads as
\beq\fl\qquad
\label{Kdef}
K={\mathcal E}(K,m)[m+\hat\nu(K,m)]
={\mathcal E}(K,m)[m+\hat\nu(K,m)^{\hat a}e(m)_{\hat a}]\,,
\eeq
where $\hat\nu(K,m)$ is a spacelike vector which identifies the local line of sight of the observer (\ref{obser}). It satisfies the unitary condition
\beq
\hat\nu(K,m)\cdot\hat\nu(K,m)=1\,,
\eeq
the dot denoting scalar product
with respect to the metric (\ref{metric_coeffs}).  To first order in $h$, then,  we have\footnote{
As it is customary, numbers in round brackets denote the order of approximation.
}
\beq
\label{vincexp}
{\mathcal E}(K,m)= 1+\epsilon^2 {\mathcal E}_{(2)}\,, \qquad
\hat\nu(K,m)^{\hat a}= n^{\hat a}+\epsilon^2 \hat\nu^{\hat a}_{(2)}\,,
\eeq
where  $n^{\hat a}$ is the unperturbed local photon direction and we have assumed the unperturbed value of  ${\mathcal E}(K,m)$ equal to unity without loss of generality.
The \lq\lq actual" locally spatial  photon direction $\hat\nu(K,m)^{\hat a}$ 
evaluated at the observation point and later denoted as $\hat\nu^{\hat a}(0)$, 
is to be considered known being related to direct observations and to the 
selected attitude of the observer's frame. This problem has been solved 
analytically in \cite{bicrofdf}.
Hence, the photon 4-momentum (\ref{Kdef}) reads, with respect to the given tetrad (\ref{obser}) and (\ref{fram}) as
\beq\fl\qquad
\label{Kexp}
K=\left[1+\epsilon^2\left({\mathcal E}_{(2)}+\frac{h}{2}\right)\right]\partial_0+\left\{n^{\hat a}+\epsilon^2 \left[\hat\nu^{\hat a}_{(2)}+n^{\hat a}\left({\mathcal E}_{(2)}-\frac{h}{2}\right)\right]\right\}\partial_a\,,
\eeq
and the perturbed orbit is given by
\beq
\label{pertorbit}
x^\alpha=x^\alpha_{(0)}+\epsilon^2x^\alpha_{(2)}\,.
\eeq
Further we shall denote as $A$ the event of observation with coordinates $x^\alpha_A=(x^0_A,x^i_A)$ and as $S$ the event of emission with coordinates $x^\alpha_S=(x^0_S,x^i_S)$.
Moreover, we fix the affine parameter $\lambda$ on the light ray trajectory so that it is $\lambda=0$ at $A$ and $\lambda=\lambda_S$ at $S$.
It is clear that while $x^\alpha_A$ are supposed to be the known position of the satellite at the observation, the coordinates $x^\alpha_S$ are our main unknowns.
The unperturbed orbit can then be written as
\beq
\label{newtorb}
x^0_{(0)}(\lambda)=\lambda+x^0_A\,, \qquad
x^i_{(0)}(\lambda)=n^i\lambda+x^i_A\,.
\eeq

Since the metric does not depend explicitly on $x^0$, the perturbed geodesic equations reduce to the system
\begin{eqnarray}
\label{perteqs1}
\frac{\rmd{\mathcal E}_{(2)}}{\rmd\lambda}&=& \frac12\frac{\rmd h}{\rmd\lambda}\,, \qquad\qquad
\frac{\rmd\hat\nu^{\hat a}_{(2)}}{\rmd\lambda} =-n^{\hat a}\frac{\rmd h}{\rmd\lambda}+\partial_ah\,,
\end{eqnarray}
which needs to be coupled to the following equations to fully determine the trajectory
\beq
\label{perteqs2}
\frac{\rmd x^0_{(2)}}{\rmd\lambda}={\mathcal E}_{(2)}+\frac12h\,, \qquad\quad
\frac{\rmd x^a_{(2)}}{\rmd\lambda}=\hat\nu^{\hat a}_{(2)}+n^{\hat a}\left({\mathcal E}_{(2)}-\frac12h\right)\,.
\eeq
Notice that the metric function $h$ and the perturbed quantities $x^0_{(2)}$, $x^a_{(2)}$ and $\hat\nu^{\hat a}_{(2)}$ all depend on the affine parameter $\lambda$ through the spatial coordinates evaluated along the world line of the photon, according to
\beq
h(x^i)\vert_{x^i=x^i(\lambda)}\equiv h(\lambda)\,,
\eeq
with an obvious abuse of notation; here $h$ plays the role of a general function of the coordinates.
Let us introduce the following quantities, which we shall use shortly
\beq\fl\quad
\label{Fdef}
H(\lambda)=\int_0^{\lambda}h(\lambda)\rmd\lambda\,,\quad
H^a(\lambda)=\int_0^{\lambda}[\partial_ah](\lambda)\rmd\lambda\,,\quad
{\mathcal H}^a(\lambda)=\int_0^{\lambda}H^a(\lambda)\rmd\lambda
\eeq
and solve the system of equations (\ref{perteqs1}) and (\ref{perteqs2}) assuming that all perturbation quantities $x^\alpha_{(2)}$, ${\mathcal E}_{(2)}$ and $\hat\nu^{\hat a}_{(2)}$ have well defined constant values at the observation point, i.e., for $\lambda=0$.
The first of Eq. (\ref{perteqs1}) gives
\beq
{\mathcal E}_{(2)}(\lambda)={\mathcal E}_{(2)}(0)+\frac12[h(\lambda)-h(0)]\,,
\eeq
while the second of Eq. (\ref{perteqs1}) can be formally integrated with solution
\beq
\label{nu2fin}
\hat\nu^{\hat a}_{(2)}(\lambda)=\hat\nu^{\hat a}_{(2)}(0)-n^{\hat a}[h(\lambda)-h(0)]+H^a(\lambda)\,.
\eeq
Eqs. (\ref{perteqs2}) thus become
\beq\fl\quad
\label{perteqs2fin}
\frac{\rmd x^0_{(2)}}{\rmd\lambda}= h(\lambda)+{\mathcal E}_{(2)}(0)-\frac12h(0)\,, \qquad
\frac{\rmd x^a_{(2)}}{\rmd\lambda}=\hat\nu^{\hat a}_{(2)}+n^{\hat a}\left({\mathcal E}_{(2)}(0)-\frac12h(0)\right)\,,
\eeq
with solution
\begin{eqnarray}
\label{pertsol2fin}
x^0_{(2)}(\lambda)&=&x^0_{(2)}(0)+H(\lambda)+\left({\mathcal E}_{(2)}(0)-\frac12h(0)\right)\lambda\,, \nonumber\\
x^a_{(2)}(\lambda)&=&x^a_{(2)}(0)-n^a[x^0_{(2)}(\lambda)-x^0_{(2)}(0)]+{\mathcal H}^a(\lambda)\nonumber\\
&&+[\hat\nu^{\hat a}_{(2)}(0)+2n^{\hat a}{\mathcal E}_{(2)}(0)]\lambda\,.
\end{eqnarray}
The photon 4-momentum (\ref{Kexp}) thus becomes
\begin{eqnarray}
\label{Kexp2}
K&=&\left[1+\epsilon^2\left(h+{\mathcal E}_{(2)}(0)-\frac{h(0)}{2}\right)\right]\partial_0\nonumber\\
&&+\left\{n^{\hat a}+\epsilon^2\left[\hat\nu^{\hat a}_{(2)}+n^{\hat a}\left({\mathcal E}_{(2)}(0)-\frac{h(0)}{2}\right)\right]\right\}\partial_a\,.
\end{eqnarray}

Noticeably most of the current literature on this topic uses the coordinate time $x^0$ instead of $\lambda$ as a parameter along the photon world line \cite{willbook,kop1,kop2,klioner1,klioner2}.
The corresponding relations are listed in \ref{timeparam} for completeness.
It should be recalled also that, irrespective to the choice of the parameter on the photon trajectory, the main task of any astrometric model is to determine $x^\alpha_S$, namely the spacetime position of the emitting star.
Finally, from the linearity of the perturbation equations, it follows that it is enough to limit the analysis to solutions involving a single body, the extension to the general case being straightforward (we plan to review this specific topic in a forthcoming paper fully dedicated to the astrometric problem in the context of Gaia mission).

\subsection{Single body}

Let us now consider as source of the background metric a single body with mass $M$ and quadrupole moment $J_2$;
 the center of mass of this body is also taken as  origin of the coordinate system.
To the order $\epsilon^2$ we then have, as stated,
\beq
\label{hdef2}
h=h_M+h_{J_2}=\frac{2GM}{r}+\frac{GMd^2}{r^3}\left(1-\frac{3z^2}{r^2}\right)J_2\,,
\eeq
where the coordinates take their unperturbed values (\ref{newtorb}), so that
\beq
r=|{\bf x}_{(0)}|
=\sqrt{\delta_{ab}x_{(0)}^ax_{(0)}^b}
=\sqrt{\lambda^2+r_A^2+2\lambda({\mathbf x}_A\cdot {\mathbf n})}\,,
\eeq
and $r_A=|{\mathbf x}_A|$.
Quantities in bold are three dimensional vectors, so that both scalar and cross product between them are meant to be the standard operations in an Euclidean space.
Similarly we have $H=H_M+H_{J_2}$, and so for the other functions $H^a$ and ${\mathcal H}^a$.
The solution (\ref{pertorbit}) for the perturbed orbit can then be written as
\beq
\label{xsoltot}
x^\alpha=x^\alpha_{(0)}+\epsilon^2x^\alpha_{(2)M}+\epsilon^2x^\alpha_{(2)J_2}\,,
\eeq
where $x^\alpha_{(2)M}$ and $x^\alpha_{(2)J_2}$ are obtained from Eq.~(\ref{pertsol2fin}) with $h=h_M$ and $h=h_{J_2}$, respectively.
Analogously, the frame components of the spatial velocity (\ref{vincexp}) which, we recall, identify the local line of sight of the observer (\ref{obser}), can be written as
\beq
\label{nusoltot}
\hat\nu(K,m)^{\hat a}= n^{\hat a}+\epsilon^2 \hat\nu^{\hat a}_{(2)M}+\epsilon^2 \hat\nu^{\hat a}_{(2)J_2}\,.
\eeq

Equations (\ref{xsoltot}) and (\ref{nusoltot}) are our main results since they provide the analytical solution to the inverse ray tracing problem in the presence of gravitational sources with not negligible mass quadrupole.

As we can see from Eqs.~(\ref{nu2fin}) and (\ref{pertsol2fin}) the perturbations $x^\alpha_{(2)}$ and $\hat\nu^{\hat a}_{(2)}$ critically depend on the functions $H(\lambda)$, $H^a(\lambda)$ and ${\mathcal H}^a(\lambda)$.
Their analytical values are given below.
The contribution of the mass monopole is given by
\begin{eqnarray}
H_M(\lambda)&=&2GM\ln\left[\frac{r+({\mathbf x}\cdot{\mathbf n})}{r_A+({\mathbf x}_A\cdot{\mathbf n})}\right]\,, \nonumber\\
H_M^a(\lambda)&=&2GM\left[n^a\left(\frac1r-\frac1{r_A}\right)-\frac{b^a}{b^2}\left(\frac{{\mathbf x}\cdot{\mathbf n}}{r}-\frac{{\mathbf x}_A\cdot{\mathbf n}}{r_A}\right)\right]\,, \nonumber\\
{\mathcal H}_M^a(\lambda)&=&n^a\left[H_M(\lambda)-\frac{2GM}{r_A}[({\mathbf x}\cdot{\mathbf n})-({\mathbf x}_A\cdot{\mathbf n})]\right]\nonumber\\
&&-\frac{2GM}{b^2}b^a\left\{r-r_A-\frac{{\mathbf x}_A\cdot{\mathbf n}}{r_A}[({\mathbf x}\cdot{\mathbf n})-({\mathbf x}_A\cdot{\mathbf n})]\right\}\,,
\end{eqnarray}
where
\beq\fl\qquad
b^a=[{\mathbf n}\times({\mathbf x}_A\times {\mathbf n})]^a=x^a_A-n^a({\mathbf x}_A\cdot{\mathbf n})\,, \qquad
b^2=r_A^2-({\mathbf x}_A\cdot{\mathbf n})^2\,,
\eeq
and
\beq
({\mathbf x}\cdot{\mathbf n})=({\mathbf x}_A\cdot{\mathbf n})+\lambda\,,
\eeq
to the order $\epsilon^2$, so that
\beq
r=\sqrt{b^2+({\mathbf x}\cdot{\mathbf n})^2}\,, \qquad
\frac{\rmd r}{\rmd\lambda}=\frac{{\mathbf x}\cdot{\mathbf n}}{r}\,.
\eeq
Finally, from Eqs.~(\ref{nu2fin})   we have
\beq
\label{solfinM}
\hat\nu^{\hat a}_{(2)M}-\hat\nu^{\hat a}_{(2)M}(0)=-\frac{2GM}{b^2}\,b^a\left(\frac{{\mathbf x}\cdot{\mathbf n}}{r}-\frac{{\mathbf x}_A\cdot{\mathbf n}}{r_A}\right)\,
\eeq
and from (\ref{pertsol2fin}) we deduce

\begin{eqnarray}\fl\quad
 \nonumber \\
\fl\quad
x^0_{(2)M}-x^0_{(2)M}(0)&=&2GM\ln\left[\frac{r+({\mathbf x}\cdot{\mathbf n})}{r_A+({\mathbf x}_A\cdot{\mathbf n})}\right]\nonumber \\
\fl\quad
&&+\left({\mathcal E}_{(2)}(0)-\frac{GM}{r_A}\right)[({\mathbf x}\cdot{\mathbf n})-({\mathbf x}_A\cdot{\mathbf n})]\,, \nonumber \\
\fl\quad
x^a_{(2)M}-x^a_{(2)M}(0)&=&-2GM\frac{b^a}{b^2}\left\{r-r_A-\frac{{\mathbf x}_A\cdot{\mathbf n}}{r_A}[({\mathbf x}\cdot{\mathbf n})-({\mathbf x}_A\cdot{\mathbf n})]\right\}\nonumber \\
&&+\left[\hat\nu^{\hat a}_{(2)M}(0)+n^a\left({\mathcal E}_{(2)}(0)-\frac{GM}{r_A}\right)\right][({\mathbf x}\cdot{\mathbf n})-({\mathbf x}_A\cdot{\mathbf n})]\,.
\end{eqnarray}
Similarly, the contribution of the quadrupole  is given by
\begin{eqnarray}\fl\quad
\label{solfinquad}
H_{J_2}(\lambda)&=&GMd^2J_2\left\{
2n_zb_z\left(\frac{1}{r^3}-\frac{1}{r_A^3}\right)
+\left(n_z^2-\frac{b_z^2}{b^2}\right)\left(\frac{{\mathbf x}\cdot{\mathbf n}}{r^3}-\frac{{\mathbf x}_A\cdot{\mathbf n}}{r_A^3}\right)\right.\nonumber\\
\fl\quad
&&\left.
+\frac1{b^2}\left(1-n_z^2-\frac{2b_z^2}{b^2}\right)\left(\frac{{\mathbf x}\cdot{\mathbf n}}{r}-\frac{{\mathbf x}_A\cdot{\mathbf n}}{r_A}\right)
\right\}\,, \nonumber \\
\fl\quad
H^a_{J_2}(\lambda)&=&GMd^2J_2\left\{
b^a\left[
-6n_zb_z\left(\frac{1}{r^5}-\frac{1}{r_A^5}\right)
-3\left(n_z^2-\frac{b_z^2}{b^2}\right)\left(\frac{{\mathbf x}\cdot{\mathbf n}}{r^5}-\frac{{\mathbf x}_A\cdot{\mathbf n}}{r_A^5}\right)
\right.\right.\nonumber\\
\fl\quad
&&\left.\left.
-\frac1{b^2}\left(1-n_z^2-\frac{4b_z^2}{b^2}\right)\left[\frac{2}{b^2}\left(\frac{{\mathbf x}\cdot{\mathbf n}}{r}-\frac{{\mathbf x}_A\cdot{\mathbf n}}{r_A}\right)
+\frac{{\mathbf x}\cdot{\mathbf n}}{r^3}-\frac{{\mathbf x}_A\cdot{\mathbf n}}{r_A^3}\right]
\right]
\right.\nonumber\\
\fl\quad
&&\left.
+n^a\left[
(1-5n_z^2)\left(\frac{1}{r^3}-\frac{1}{r_A^3}\right)
+3b^2\left(n_z^2-\frac{b_z^2}{b^2}\right)\left(\frac{1}{r^5}-\frac{1}{r_A^5}\right)\right.\right.\nonumber\\
\fl\quad
&&\left.\left.
+\frac{2n_zb_z}{b^2}\left[\frac{2}{b^2}\left(\frac{{\mathbf x}\cdot{\mathbf n}}{r}-\frac{{\mathbf x}_A\cdot{\mathbf n}}{r_A}\right)
+\frac{{\mathbf x}\cdot{\mathbf n}}{r^3}-\frac{{\mathbf x}_A\cdot{\mathbf n}}{r_A^3}\right]
\right.\right.\nonumber\\
\fl\quad
&&\left.\left.
-6n_zb_z\left(\frac{{\mathbf x}\cdot{\mathbf n}}{r^5}-\frac{{\mathbf x}_A\cdot{\mathbf n}}{r_A^5}\right)
\right]
\right\}\,, \nonumber\\
\fl\quad
{\mathcal H}^a_{J_2}(\lambda)&=&GMd^2J_2\left\{
b^a\left[
\left(n_z^2-\frac{b_z^2}{b^2}\right)\left(\frac{1}{r^3}-\frac{1}{r_A^3}\right)
\right.\right.\nonumber\\
\fl\quad
&&\left.\left.
-\frac{2n_zb_z}{b^2}\left[\frac{2}{b^2}\left(\frac{{\mathbf x}\cdot{\mathbf n}}{r}-\frac{{\mathbf x}_A\cdot{\mathbf n}}{r_A}\right)
+\frac{{\mathbf x}\cdot{\mathbf n}}{r^3}-\frac{{\mathbf x}_A\cdot{\mathbf n}}{r_A^3}\right]
\right.\right.\nonumber\\
\fl\quad
&&\left.\left.
+\frac1{b^2}\left(1-n_z^2-\frac{4b_z^2}{b^2}\right)\left[-\frac{2}{b^2}(r-r_A)+\frac{1}{r}-\frac{1}{r_A}\right]
\right.\right.\nonumber\\
\fl\quad
&&\left.\left.
+\left[
\frac1{b^2}\left(1-n_z^2-\frac{4b_z^2}{b^2}\right)\left(\frac{2}{b^2}+\frac{1}{r_A^2}\right)\frac{{\mathbf x}_A\cdot{\mathbf n}}{r_A}
+3\left(n_z^2-\frac{b_z^2}{b^2}\right)\frac{{\mathbf x}_A\cdot{\mathbf n}}{r_A^5}\right.\right.\right.\nonumber\\
\fl\quad
&&\left.\left.\left.
+\frac{6n_zb_z}{r_A^5}\right][({\mathbf x}\cdot{\mathbf n})-({\mathbf x}_A\cdot{\mathbf n})]
\right]\right.\nonumber\\
\fl\quad
&&\left.
+n^a\left[
-\frac{2n_zb_z}{b^2}\left[-\frac{2}{b^2}(r-r_A)+\frac{1}{r}-\frac{1}{r_A}\right]
-\frac{2n_z^2}{b^2}\left(\frac{{\mathbf x}\cdot{\mathbf n}}{r}-\frac{{\mathbf x}_A\cdot{\mathbf n}}{r_A}\right)
\right.\right.\nonumber\\
\fl\quad
&&\left.\left.
+\left[
\frac{2n_z^2}{r_A^3}-\frac{2n_zb_z}{b^2}\left(\frac{2}{b^2}+\frac{1}{r_A^2}\right)\frac{{\mathbf x}_A\cdot{\mathbf n}}{r_A}
\right.\right.\right.\nonumber\\
\fl\quad
&&\left.\left.\left.
-\frac{1}{r_A^3}\left(1-\frac{3z_A^2}{r_A^2}\right)\right][({\mathbf x}\cdot{\mathbf n})-({\mathbf x}_A\cdot{\mathbf n})]
+H_{J_2}(\lambda)
\right]
\right\}\,.
\end{eqnarray}
Finally, the solution for $\hat\nu^{\hat a}_{(2)J_2}$ and $x^\alpha_{(2)J_2}$ are obtained from Eqs.~(\ref{nu2fin}) and (\ref{pertsol2fin}), respectively.

As stated, the above corrections provide the analytical solutions (\ref{xsoltot}) and (\ref{nusoltot}) which are basic to multipolar relativistic astrometry.  For instance, this explicit solution of the photon equation will allow a complete weak field analysis of the gravitational lensing by objects endowed with nonzero quadrupole moment.
The necessity of having quadrupolar corrections to light deflection formula has been largely addressed in the literature (see, e.g., Refs. \cite{Dam-EF,kop4} and references therein), but only here the analysis has been performed by comparing the case of an exact solution of Einstein's equations with its weak field limit in a unified treatment. Such corrections have been also discussed in connection with gravitational lensing measurements (see Ref. \cite{ruth}and references therein). 
Furthermore, numerical estimates for observational quantities related to this approximate solution for most of Solar System objects have been studied by several authors (see, e.g., Refs. \cite{cromi,kop3,klioner1,klioner2,klioner4}).
The results of the present section are in agreement with the existing literature (see also \ref{timeparam}, where the relation with different parametrizations, coordinates and conventions adopted in other approaches is discussed too), so that we will not repeat such a discussion here.

\subsection{RAMOD master equation: notations in comparison}

Relativistic Astrometric MODels (RAMOD) have been developed by de Felice and coworkers over a period of about ten years to theoretically support the astrometric mission Gaia. It is most useful, then, to relate the notation of the present paper (recently standardized in Ref. \cite{fdfbini}) to the corresponding one used in the series of RAMOD papers \cite{ramod1,ramod2,ramodx,ramody,ramod3,ramod4,cro2011}, where the analysis of ray tracing has been performed in the general context of an unspecified perturbative solution of the Einstein's field equations, i.e., with special attentions to the equations associated with tracing and their initial conditions.
Differently, the approximate solution describing the gravitational field of a massive body endowed with quadrupole moment at rest at the origin of the coordinates has been adopted here.

RAMOD master equation, in the static case, is written in the form (see Ref. \cite{ramod3}, Eq. (19), slightly manipulated by replacing $h_{00}=h$, $h_{ij}=h\delta_{ij}$ according to Eqs. (\ref{metric_coeffs}) of the present paper)
\beq
\label{RAMODme}
\frac{d\bar \ell^k}{\rmd \sigma}=-\frac32 \bar \ell^k\bar \ell^i \partial_i h + \partial_k h\,,
\eeq
where $h$, as in this case, is the perturbation to the Minkowski metric. Such equation coincides with our Eq. (\ref{geo_gen}) (and its further manipulation (\ref{perteqs1})) if one denotes
\beq
\bar \ell =\hat \nu (K,m)\,,
\eeq
and uses frame components $\hat \nu^{\hat a}$ with respect to the frame (\ref{fram}) instead of coordinate components $\hat \nu^a$, which are related by $\hat \nu^{\hat a}=[1+\epsilon^2(h/2)]\hat \nu^a$.
RAMOD decomposition of the photon momentum is in fact
\beq
K=-(m\cdot K)m+\ell= {\mathcal E}(K,m)m+ \ell\,,
\eeq
so that we have also $\ell ={\mathcal E}(K,m)\hat \nu (K,m) $. Finally, the affine parameter $\sigma$ coincides exactly with $\lambda$ used in this paper (see also \cite{cro2011}).

A detailed discussion of the actual observation equation  associated with Eq. (\ref{RAMODme}) and adapted to the Gaia 
mission is out of scope here and then it will be treated in a forthcoming paper fully dedicated to this purpose.
However, we can outline the principles of the procedure, which are based on the fact that the observables can be expressed as functions of the direction cosines of the incoming direction with respect to the axes of a tetrad adapted to the Gaia satellite attitude, say $E_{\hat a}^\alpha$ (see Refs. \cite{crovec,biavebu} for details).
Let $\cos\psi_{({\hat a},{K})}=\hat \nu(K,m) \cdot E_{\hat a}$ be the $\hat a$-th direction cosine measured by the observer on the satellite.
This latter expression can then be linked in a standard way to the usual astrometric unknowns  of the observed object (position, like the equatorial coordinates, their corresponding proper motions, and the parallax).

\section{Concluding remarks}

In this paper we have investigated some optical properties of the Erez-Rosen metric, which is an exact solution of Einstein equations describing the spacetime of a static body endowed with a mass quadrupole moment.

We have considered first the curvature invariants and have deduced some properties of the solution which clearly show the effects of the quadrupole on the gravitational strength of the source.
We have then analyzed the geodesic properties of the solution focusing on null orbits in the symmetry plane.
As a result we have found the pattern of the light trajectories close to the source as a function of the quadrupole parameter, which specifies the shape of the body. An interesting outcome of this analysis is the existence of spatially bound photon orbits close to the source due to the quadrupole action.
Pursuing in our study, we have tackled the problem of the gravitational lensing induced by the source and have obtained explicit expressions for standard lensing observables, like image positions and magnification, in the limit of weak field and small quadrupole, generalizing previous results for static spherically symmetric bodies.

The last part of the paper is devoted to the analytical solution of the inverse ray tracing problem in the post-Newtonian approximation of the Erez-Rosen metric.
We have found that, in this limit, the Erez-Rosen metric reduces to the metric form used in standard astrometric modeling, like RAMOD. 
Hence our analysis, which allows to include in those models the quadrupole corrections, is fully consistent with a general relativistic approach.
Furthermore, we have shown that this result is also consistent with other approaches, in spite of the use of different conventions, parameters and coordinates. 

In particular, in order to fully exploit the Erez Rosen solution also
in the context of relativistic astrometry,  we have  given an analytical
solution in the case of a static massive body with quadrupole corrections
which is directly connected  to the standard approach used in RAMOD. 
This result is particularly relevant for the interpretation of high precision measurements as expected from the ESA mission Gaia to be started in 2013.

\section*{Acknowledgments}
M.C. and A.V. acknowledge the Italian Space Agency (ASI) for the support given under the contract to INAF I/058/10/0 (Gaia Mission - The Italian Participation to DPAC).
All  authors are indebted to the unknown referees for their valuable work in raising few questions, the answer to which has much improved the paper.

\appendix

\section{Connection between the Erez-Rosen solution and the Hartle-Thorne solution}
\label{ERtoHT}

Consider the general static line element (\ref{metric_Weyl}) with metric functions $f=e^{2\psi}$ and $\gamma$.
From a given solution $(\psi,\gamma)$, a set of new solutions may be generated by the transformation
\beq
\label{ZVtrasf}
\psi\to\delta\psi\,, \qquad
\gamma\to\delta^2\gamma\,,
\eeq 
where $\delta$ is a real number corresponding to the Zipoy-Voorhees \cite{zipoy,voorhees} parameter.
For our purposes it is convenient to set such a parameter as $\delta=1+sq$, where $s$ is a real number.
By applying the Zipoy-Voorhees transformation (\ref{ZVtrasf}) to the Erez-Rosen solution (\ref{metdef}) one obtains a new set of solutions to linear order in $q$ with metric functions 
\begin{eqnarray}
f&\simeq&\frac{x-1}{x+1}\left[1-q\left(2P_2Q_2-s\ln\frac{x-1}{x+1}\right)\right]\,, \nonumber\\
\gamma&\simeq&\frac12[1+2(1+s)q]\ln\frac{x^2-1}{x^2-y^2} + 2q(1-P_2)Q_1\,.
\end{eqnarray}
For $s=0$, i.e. $\delta=1$, we recover the linearized form of the Erez-Rosen solution (\ref{metdef}).
It also contains the Hartle-Thorne solution (in the absence of rotation) for $s=-1$.
To show this, introduce first standard Schwarzschild-like coordinates through the transformation (\ref{trasftoBL}), i.e., $x={r}/{M}-1$ and $y=\cos\theta$, with $t$ and $\phi$ unchanged.
Then set $s=-1$ and $M={\mathcal M}(1+q)$.
The further transformation $r=r(R,\Theta)$ and $\theta=\theta(R,\Theta)$ with 
\begin{eqnarray}\fl\quad
r&=&R+{\mathcal M}q+\frac32{\mathcal M}q\sin^2\Theta\left[\frac{R}{\mathcal M}-1+\frac12\frac{R^2}{{\mathcal M}^2}\left(1-\frac{2{\mathcal M}}{R}\right)\ln\left(1-\frac{2{\mathcal M}}{R}\right)\right]\,, \nonumber\\
\fl\quad
\theta&=&\Theta-\frac32q\sin\Theta\cos\Theta\left[2+\left(\frac{R}{\mathcal M}-1\right)\ln\left(1-\frac{2{\mathcal M}}{R}\right)\right]
\end{eqnarray}
finally gives the mapping between the general form of the Erez-Rosen solution and the Hartle-Thorne metric (\ref{HTmet}) with mass and quadupole moment given by 
\beq
{\mathcal M}=M(1-q)\,, \qquad 
{\mathcal Q}=\frac45M^3q\,,
\eeq
respectively.
Further details on such a derivation can be found in the Appendix B of Ref. \cite{mt91}.

\section{Time parametrization of the photon orbit}
\label{timeparam}

If one uses as parameter on the trajectory the time coordinate $x^0$ instead of $\lambda$, the photon 4-momentum (\ref{Kdef}) can be written as
\beq
K=\partial_0+[n^{\hat a}+\epsilon^2 (\hat\nu^{\hat a}_{(2)}-n^{\hat a}h)]\partial_a\,.
\eeq
Taking the derivative of Eq. (\ref{pertorbit}) with respect to $\lambda$ yields
\begin{eqnarray}
\frac{\rmd x^0}{\rmd\lambda}&=&1+\epsilon^2\left(h(\lambda)+{\mathcal E}_{(2)}(0)-\frac12h(0)\right)\,,\nonumber\\
\frac{\rmd x^a}{\rmd\lambda}&=&n^{\hat a}+\epsilon^2 \left[\hat\nu^{\hat a}_{(2)}(\lambda)+n^{\hat a}\left({\mathcal E}_{(2)}(0)-\frac12h(0)\right)\right]\,,
\end{eqnarray}
so that eliminating $\lambda$ implies
\beq
\label{eqxadit}
\frac{\rmd x^a}{\rmd x^0}
=n^{\hat a}+\epsilon^2(\hat\nu^{\hat a}_{(2)}-n^{\hat a}h)\,,
\eeq
where the dependence on time on the right hand side is implicit through the spatial coordinates.
Once differentiated with respect to $x^0$, the previous equation then gives
\beq
\label{eqxadit2ord}
\frac{\rmd^2 x^a}{\rmd (x^0)^2}=\epsilon^2\left(\frac{\rmd \nu^{\hat a}_{(2)}}{\rmd x^0}-n^{\hat a}\frac{\rmd h}{\rmd x^0}\right)
=\epsilon^2\left(\partial_ah-2n^{\hat a}\frac{\rmd h}{\rmd x^0}\right)\,.
\eeq
This is the form of the equations describing the corrections to light propagation in a barycentric coordinate system usually adopted in the literature (see, e.g., Ref. \cite{willbook}).
The corresponding solution can in turn be written as
\beq
\label{soltot}
x^a=x^a_S+n^a(x^0-x^0_S)+\epsilon^2x^a_{(2)M}+\epsilon^2x^a_{(2)J_2}\,,
\eeq
by separating the contributions due to the monopole and the quadrupole.

In order to compute the total light deflection of a photon emitted at time $t_S$ and spatial coordinates $x^a(t_S)$ it is enough to solve the associated initial value problem \cite{klioner2}
\beq
x^a_S=x^a(t_S)\,,\qquad
n^a=\lim_{x^0\to-\infty}\frac{\rmd x^a}{\rmd x^0}\,.
\eeq
According to Eqs. (\ref{eqxadit2ord}) and (\ref{soltot}), the initial conditions for the relativistic corrections are
\beq\fl\qquad
x^a_M(x^0_S)=0=x^a_{J_2}(x^0_S)\,, \qquad
\lim_{x^0\to-\infty}\frac{\rmd x^a_M}{\rmd x^0}=0=\lim_{x^0\to-\infty}\frac{\rmd x^a_{J_2}}{\rmd x^0}\,.
\eeq
Let $\Sigma^a$ denote the limiting asymptotic values of the coordinate components of the unit tangent vector to the light path for $x^0\to+\infty$.
The total light deflection is then given by $|{\bf n}\times{\bf\Sigma}|$.

However, for practical modeling of observations of Solar System objects it is not sufficient to consider the initial value problem for light propagation.
Hence, one has to solve the boundary value problem \cite{klioner1,theyssa}
\beq
x^a_S=x^a(t_S)\,,\qquad
x^a_A=x^a(t_A)\,,
\eeq
the light ray being emitted at time $t_S$ and spatial position $x^a(t_S)$ and then received at time $t_A$ and spatial position $x^a(t_A)$.
In this case one can compute for instance the time of light propagation for the given boundary conditions.

For a single body, following the same procedure as in Section 3, the monopole solution to the initial value problem turns out to be
\begin{eqnarray}
\label{solxaditM}
x^a_{(2)M}&=&-2GM\left\{
n^a\ln\left[\frac{r+({\mathbf x}\cdot{\mathbf n})}{r_S+({\mathbf x}_S\cdot{\mathbf n})}\right]
\right.\nonumber\\
&&\left.
+\frac{b^a}{b^2}\left[r-r_S-\frac{{\mathbf x}_S\cdot{\mathbf n}}{r_S}[({\mathbf x}\cdot{\mathbf n})-({\mathbf x}_S\cdot{\mathbf n})]\right]
\right\}\,,
\end{eqnarray}
where now
\beq
r=\sqrt{(x^0-x^0_S)^2+r_S^2+2(x^0-x^0_S)({\mathbf x}_S\cdot{\mathbf n})}\,,
\eeq
with $r_S=|{\mathbf x}_S|$ and
\beq
({\mathbf x}\cdot{\mathbf n})=({\mathbf x}_S\cdot{\mathbf n})+x^0-x^0_S\,,
\eeq
to the order $\epsilon^2$.
The monopole correction to the frame components of the velocity is
\beq
\nu^{\hat a}_{(2)M}=-\frac{2GM}{b^2}\,b^a\left(\frac{{\mathbf x}\cdot{\mathbf n}}{r}-\frac{{\mathbf x}_S\cdot{\mathbf n}}{r_S}\right)\,.
\eeq
The solution for the quadrupole contribution $x^a_{J_2}$ can be also obtained straightforwardly
\begin{eqnarray}\fl\quad
x^a_{(2)J_2}&=&GMd^2J_2\left\{
b^a\left[
\left(n_z^2-\frac{b_z^2}{b^2}\right)\left(\frac{1}{r^3}-\frac{1}{r_S^3}\right)
\right.\right.\nonumber\\
\fl\quad
&&\left.\left.
-\frac{2n_zb_z}{b^2}\left[\frac{2}{b^2}\left(\frac{{\mathbf x}\cdot{\mathbf n}}{r}-\frac{{\mathbf x}_S\cdot{\mathbf n}}{r_S}\right)
+\frac{{\mathbf x}\cdot{\mathbf n}}{r^3}-\frac{{\mathbf x}_S\cdot{\mathbf n}}{r_S^3}\right]
\right.\right.\nonumber\\
\fl\quad
&&\left.\left.
+\frac1{b^2}\left(1-n_z^2-\frac{4b_z^2}{b^2}\right)\left[-\frac{2}{b^2}(r-r_S)+\frac{1}{r}-\frac{1}{r_S}\right]
\right.\right.\nonumber\\
\fl\quad
&&\left.\left.
+\left[
\frac1{b^2}\left(1-n_z^2-\frac{4b_z^2}{b^2}\right)\left(\frac{2}{b^2}+\frac{1}{r_S^2}\right)\frac{{\mathbf x}_S\cdot{\mathbf n}}{r_S}
+3\left(n_z^2-\frac{b_z^2}{b^2}\right)\frac{{\mathbf x}_S\cdot{\mathbf n}}{r_S^5}\right.\right.\right.\nonumber\\
\fl\quad
&&\left.\left.\left.
+\frac{6n_zb_z}{r_S^5}\right][({\mathbf x}\cdot{\mathbf n})-({\mathbf x}_S\cdot{\mathbf n})]
\right]\right.\nonumber\\
\fl\quad
&&\left.
+n^a\left[
-\frac{2n_zb_z}{b^2}\left[-\frac{2}{b^2}(r-r_S)+\frac{1}{r}-\frac{1}{r_S}+b^2\left(\frac{1}{r^3}-\frac{1}{r_S^3}\right)\right]
\right.\right.\nonumber\\
\fl\quad
&&\left.\left.
-\frac1{b^2}\left(1+n_z^2-\frac{2b_z^2}{b^2}\right)\left(\frac{{\mathbf x}\cdot{\mathbf n}}{r}-\frac{{\mathbf x}_S\cdot{\mathbf n}}{r_S}\right)
-\left(n_z^2-\frac{b_z^2}{b^2}\right)\left(\frac{{\mathbf x}\cdot{\mathbf n}}{r^3}-\frac{{\mathbf x}_S\cdot{\mathbf n}}{r_S^3}\right)
\right.\right.\nonumber\\
\fl\quad
&&\left.\left.
+\left[
\frac{2n_z^2}{r_S^3}-\frac{2n_zb_z}{b^2}\left(\frac{2}{b^2}+\frac{1}{r_S^2}\right)\frac{{\mathbf x}_S\cdot{\mathbf n}}{r_S}
\right]
[({\mathbf x}\cdot{\mathbf n})-({\mathbf x}_S\cdot{\mathbf n})]
\right]
\right\}\,.
\end{eqnarray}
whereas the correction to the frame components of the velocity is
\begin{eqnarray}\fl\quad
\nu^{\hat a}_{(2)J_2}&=&GMd^2J_2\left\{
b^a\left[
-6n_zb_z\left(\frac{1}{r^5}-\frac{1}{r_S^5}\right)\right.\right.\nonumber\\
\fl\quad
&&\left.\left.
-\frac1{b^2}\left(1-n_z^2-\frac{4b_z^2}{b^2}\right)\left[\frac{2}{b^2}\left(\frac{{\mathbf x}\cdot{\mathbf n}}{r}-\frac{{\mathbf x}_S\cdot{\mathbf n}}{r_S}\right)
+\frac{{\mathbf x}\cdot{\mathbf n}}{r^3}-\frac{{\mathbf x}_S\cdot{\mathbf n}}{r_S^3}\right]
\right.\right.\nonumber\\
\fl\quad
&&\left.\left.
-3\left(n_z^2-\frac{b_z^2}{b^2}\right)\left(\frac{{\mathbf x}\cdot{\mathbf n}}{r^5}-\frac{{\mathbf x}_S\cdot{\mathbf n}}{r_S^5}\right)
\right]\right.\nonumber\\
\fl\quad
&&\left.
+n^a\left[
-2n_z^2\left(\frac{1}{r^3}-\frac{1}{r_S^3}\right)\right.\right.\nonumber\\
\fl\quad
&&\left.\left.
+\frac{2n_zb_z}{b^2}\left[\frac{2}{b^2}\left(\frac{{\mathbf x}\cdot{\mathbf n}}{r}-\frac{{\mathbf x}_S\cdot{\mathbf n}}{r_S}\right)
+\frac{{\mathbf x}\cdot{\mathbf n}}{r^3}-\frac{{\mathbf x}_S\cdot{\mathbf n}}{r_S^3}\right]
\right]
\right\}\,.
\end{eqnarray}

\end{document}